\documentclass[lettersize,journal]{IEEEtran}
\usepackage{amsmath,amsfonts}
\usepackage{amssymb}
\usepackage{algorithm}
\usepackage{algpseudocode}
\usepackage{array}

\usepackage{textcomp}
\usepackage{stfloats}
\usepackage{url}
\usepackage{verbatim}
\usepackage{graphicx}
\usepackage{cite}
\usepackage{ntheorem}
\usepackage{cuted}
\usepackage{xcolor}
\usepackage{diagbox}

\usepackage[font=footnotesize,labelfont=rm,textfont=rm,labelsep=period]{caption}

\usepackage{subfig}

\begin{document}

\title{Movement-Cost-Aware Energy Minimization in Pinching Antenna-Assisted WPCN: A Joint Energy–Time Optimization}

\author{
	Xiang Zhang,~\IEEEmembership{Student Member,~IEEE,}
	Ke Xiong,~\IEEEmembership{Member,~IEEE,}
	Wei Chen,~\IEEEmembership{Senior Member,~IEEE,}\\
	Pingyi Fan,~\IEEEmembership{Senior Member,~IEEE,}
	 Shenghui Song,~\IEEEmembership{Senior Member,~IEEE}
	\vspace{-20pt}

	\thanks{Xiang Zhang and Ke Xiong are with the Engineering Research Center of Network Management Technology for High Speed Railway of Ministry of Education, School of Computer Science and Technology, the Collaborative Innovation Center of Railway Traffic Safety, and the National Engineering Research Center of Advanced Network Technologies, Beijing Jiaotong University, Beijing 100044, China (e-mail: xiangzhang@bjtu.edu.cn; kxiong@bjtu.edu.cn).}
	
	\thanks{Wei Chen and Pingyi Fan are with the Department of Electronic Engineering, and also with the Beijing National Research Center for Information Science and Technology, Tsinghua University, Beijing 100084, China (e-mail: wchen@tsinghua.edu.cn; fpy@tsinghua.edu.cn).}
	
	\thanks{Shenghui Song is with the Department of Electronic and Computer Engineering, The Hong Kong University of Science and Technology, Hong Kong SAR (e-mail: eeshsong@ust.hk).}
}

\maketitle

\begin{abstract}
	This paper investigates a pinching antenna (PA)-assisted wireless powered communication network (WPCN), where the PAs are capable of moving to designated positions to construct favorable channels to facilitate wireless power transfer (WPT) and wireless information transmission (WIT). To achieve a green PA-assisted WPCN system design, a total system energy minimization problem is formulated through joint energy–time optimization, where the PA positions, PAM duration, WPT duration, WIT duration, and transmit powers of the hybrid access point (HAP) and Internet of Things devices (IoTDs) are jointly optimized, subject to the WPT and WIT quality-of-service (QoS) constraints. Different from previous works, to be practical, this work takes the PA movement (PAM) costs, including energy consumption and time overhead, into account, thus introducing a fundamental joint energy and time tradeoff between PAM and WPCN. As the resulting optimization problem is non-convex due to the multiplicative coupling between PA positions and transmit powers, it motivates the decomposition into two subproblems. However, the competition between the PAM and WPCN durations and the lack of incentive for moving PAs to reduce WPT energy consumption make the two subproblems difficult to optimize independently. To tackle it, an alternating direction method of multipliers (ADMM)-based algorithm is proposed to coordinate the coupled subproblems. Furthermore, for the special case where PAM duration is fixed, a lower-complexity alternating optimization (AO)-based algorithm is designed. Numerical results demonstrate that the presented ADMM-based and AO-based algorithms achieve near-optimal performance compared to the exhaustive search method. Moreover, compared with conventional designs, our presented movement-cost-aware PA-assisted WPCN notably reduces total energy consumption by avoiding excessive PAM, particularly under low QoS requirements.
\end{abstract}

\vspace{-10pt}
\section{Introduction}
\subsection{Background}
\IEEEPARstart{W}{ireless} powered communication networks (WPCNs) have emerged as a promising wireless communication paradigm that delivers far-field radio frequency (RF) energy to energy-constrained Internet of Things devices (IoTDs), thereby eliminating or reducing reliance on conventional batteries and wired power supplies and enabling massive IoTDs to operate sustainably over extended periods \cite{X_Zhang_2026_1, X_Zhang_2026_2}. A typical WPCN consists of two phases \cite{Y_Ge_2024}, i.e., wireless power transfer (WPT) and wireless information transmission (WIT). In the WPT phase, a hybrid access point (HAP) transfers RF energy to IoTDs, while in the WIT phase, the IoTDs transmit information to the HAP using the harvested energy. Nevertheless, the performance of WPCNs is fundamentally limited by RF propagation distance from HAP to IoTDs \cite{Y_Lu_2026}. That is, a long propagation distance leads to severe large-scale path loss, which significantly reduces the received signal power, especially for IoTDs located far from the HAP. Consequently, IoTDs may experience both insufficient harvested energy and unfavorable wireless channels, which thus restricts the achievable throughput of the information transmission.

To improve the performance of WPCNs, recent studies have explored a range of advanced antenna technologies, including fluid antennas \cite{X_Lin_2026}, movable antennas \cite{Y_Gao_2026}, and reconfigurable intelligent surfaces \cite{K_Cao_2026}, which are able to dynamically construct good-quality wireless channels to enhance WPT and WIT. In addition to the aforementioned antenna technologies, pinching antenna (PA) technology has also emerged as an effective solution capable of enabling flexible wireless channel reconfiguration \cite{Z_Yang_2026}. Specifically, PAs can be flexibly repositioned along the dielectric waveguide, allowing the radiation points to be placed at favorable locations relative to the served devices \cite{Y_Liu_2026_1}. Such spatial reconfiguration can effectively improve the channel conditions by reducing the propagation distance and establishing favorable line-of-sight (LoS) links. 

\vspace{-5pt}
\subsection{Related Work}
So far, benefiting from its advantages, PA has attracted considerable attention in the design of various wireless communication systems \cite{Y_Xu_2025, Z_Zhang_2025, Y_Ai_2026, Z_Wang_2025, Y_Xu_2026_1, V_K_Papanikolaou_2025, Y_Li_2026_1, S_Pakravan_2026, H_Hu_2026, C_Gao_2026, Z_Zhu_2026_1, K_Cao_2025, Z_Zhu_2026_2, P_Liu_2025, P_Liu_2026, Y_Li_2026_2, D_Gan_2026}. For instance, in \cite{Y_Xu_2025}, PAs were employed to enhance the transmission rate in a downlink transmission system. In \cite{Z_Zhang_2025}, PAs were deployed to establish reliable line-of-sight links in an integrated sensing and communication system. In \cite{Y_Ai_2026}, PAs were leveraged to reduce the maximum task delay in a mobile edge computing system. Particularly, owing to its capability of enhancing the efficiency of both WPT and WIT through wireless channel reconfiguration, PA technology has been introduced into WPCNs \cite{V_K_Papanikolaou_2025, Y_Li_2026_1, S_Pakravan_2026, H_Hu_2026, C_Gao_2026, Z_Zhu_2026_1, K_Cao_2025, Z_Zhu_2026_2, P_Liu_2025, P_Liu_2026, Y_Li_2026_2, D_Gan_2026}. In \cite{V_K_Papanikolaou_2025}, the authors demonstrated the potential of PA technology for addressing the doubly-near-far problem in WPCNs. In \cite{Y_Li_2026_1, S_Pakravan_2026, H_Hu_2026}, the authors respectively maximized the sum-rate, maximized the system energy efficiency, and minimized the average age of information (AoI) in the single-waveguide PA-aided WPCNs. In \cite{C_Gao_2026} and \cite{Z_Zhu_2026_1}, the authors respectively maximized the minimum uplink rate and the sum rate in the multi-waveguide PA-aided WPCNs. In \cite{K_Cao_2025} and \cite{Z_Zhu_2026_2}, the authors respectively derived the closed-form expressions of outage probability and ergodic rate for the practical lossy waveguide case and ideal lossless waveguide case, and maximized the sum throughput of users in the PA-aided WPCNs with separate energy-transmitting and information-receiving base stations. In \cite{P_Liu_2025} and \cite{P_Liu_2026}, the authors respectively maximized the computation capacity of non-orthogonal multiple access (NOMA)-enabled and time division multiple access (TDMA)-enabled PA-aided wireless-powered mobile edge computing (MEC) networks. In \cite{Y_Li_2026_2} and \cite{D_Gan_2026}, the authors respectively maximized the total received power of energy users and the sum rate of information users in the PA-aided simultaneous wireless information and power transfer (SWIPT) systems.

Nevertheless, all the aforementioned works only treated PA movement (PAM) as an ideal operation, i.e., omitting its associated costs, including energy consumption and time overhead. Actually, in the PA-assisted WPCNs, the PAs are deployed along waveguides with lengths on the order of ten meters, and their movement is realized via mechanical actuators, such as stepper motors, and inevitably incurs non-negligible energy consumption and time overhead \cite{Y_Xu_2026_2, X_Zheng_2026, J_Ding_2026, M_Liu_2026}. Therefore, omitting the energy and time costs of PAM may lead to inefficient PA-assisted WPCN system designs.

\vspace{-5pt}
\subsection{Motivations and Contributions}
{\it{To bridge the gap, we incorporate both the energy consumption and time overhead of PAM into the optimal design of PA-assisted WPCNs in this paper.}} We consider a typical PA-assisted WPCN, where multiple PAs first move along a waveguide to their designated positions, then the HAP transfers RF energy to the IoTDs through the PAs, and the IoTDs subsequently utilize the harvested energy to transmit their information to the PAs. We jointly optimize the PA positions, PAM duration, WPT duration, WIT duration, and transmit powers of HAP and IoTDs to minimize the total system energy consumption, i.e., the sum of the PAM and WPCN energy consumption, while satisfying the quality-of-service (QoS) requirement of WPCN. 

Since a tradeoff exists between the performance gains of PAM and its associated time and energy costs, optimally solving the formulated problem faces the following unique difficulty. That is, the PA positions are severely coupled with the transmit powers of the HAP and IoTDs through the wireless channels. To handle this coupling, PA position optimization and WPCN resource allocation have to be decoupled and optimized separately. However, such a decomposition introduces two additional challenges. {\it{First}}, the PAM duration competes with the WPCN duration under the total transmission block constraint, such that fixing the WPCN duration would restrict the optimization of the PAM duration. {\it{Second}}, fixing the HAP transmit power during PA position optimization may eliminate the incentive for PAM, as PAM cannot reduce the WPT energy consumption. Consequently, the two subproblems cannot be simply optimized independently. To tackle these challenges, we propose a joint energy–time optimization, which encompasses an alternating direction method of multipliers (ADMM)-based algorithm for the general case and a lower-complexity alternating optimization (AO)-based algorithm for the special case. The major contributions of this work are outlined below.

{\it{First}}, to achieve a green PA-assisted WPCN design, we formulate a total system energy minimization problem that jointly optimizes the PA positions, PAM duration, WPT duration, WIT duration, and transmit powers of HAP and IoTDs, while satisfying the QoS requirement, where energy consumption and time overhead of PAM are taken into account.

{\it{Second}}, to handle this nonconvex problem, we first decouple it into two subproblems, i.e., PA position optimization subproblem and WPCN resource allocation subproblem. To prevent the optimization of the PAM duration from being restricted by a fixed WPCN duration, an auxiliary variable is introduced to duplicate the PAM duration, and a consensus constraint is imposed to ensure that the two copies of the PAM duration converge to the same value. To preserve the incentive for PAM during PA position optimization, the HAP transmit power is introduced as an auxiliary optimization variable in the PA position optimization. The two subproblems are then alternately optimized under the augmented Lagrangian framework. This leads to an ADMM-based algorithm.

{\it{Third}}, since the PA positions affect both the large-scale path loss and the small-scale phase shift of the wireless channel, a fine-tuning-based successive convex approximation (SCA) algorithm is proposed to solve the resulting nonconvex PA position optimization subproblem. Specifically, with phase alignment assumed, the PA positions are optimized via SCA to improve the large-scale channel gains. Then, a fine-tuning mechanism is incorporated within the SCA framework to maintain phase alignment and further refine the PA positions.

{\it{Fourth}}, for the special case where the PAM duration is fixed, the PAM and WPCN durations no longer compete with each other, and thus the auxiliary variable for duplicating the PAM duration is no longer required. This leads to an AO-based algorithm with lower computational complexity.

Simulation results demonstrate that the presented algorithms achieve near-optimal performance, with optimality gaps of only $0.94\%$ and $0.87\%$ for the ADMM- and AO-based algorithms, respectively, compared with the exhaustive search (ES) method. Moreover, compared with conventional designs without taking PAM costs into account, our presented movement-cost-aware PA-assisted WPCN notably reduces total energy consumption by balancing the benefits of PAM against its associated costs, particularly under low QoS requirements, achieving energy-saving gains of up to 77.51\% and 56.61\% in the two cases, respectively. Finally, the impacts of key system parameters on the total energy consumption are quantitatively discussed.

The rest of this paper is structured as follows. Section II describes the considered system model. Section III presents the problem formulation for minimizing the total system energy consumption. Section IV presents the solution methodology for the FPD case, followed by the APD case in Section V. Section VI reports the simulation results, and Section VII concludes the paper.

\vspace{-5pt}
\section{System Model}
As shown in Fig. 1, we consider a typical PA-assisted WPCN, which consists of a HAP equipped with a waveguide supporting $M$ movable PAs to serve $U$ single-antenna IoTDs. Let $\mathcal{M}\triangleq\{1,\ldots,M\}$ and $\mathcal{U}\triangleq\{1,\ldots,U\}$ represent the sets of the PAs and IoTDs. As shown in Fig. 2, the system operates over a time block of duration $\tau$, which is divided into three consecutive phases with durations $\tau_0$, $\tau_1$, and $\tau_2$, corresponding to the PAM, WPT, and WIT phases, respectively. Specifically, during the PAM phase, the PAs are repositioned along the waveguide. Then, during the WPT phase, the HAP transfers RF energy to the IoTDs through the waveguide with all PAs radiating RF signals. Finally, the IoTDs utilize the harvested RF energy to transmit their information to the HAP during the WIT phase.

For clarity, a three-dimensional Cartesian coordinate system is established to characterize the system geometry, as illustrated in Fig. 1. Specifically, the waveguide of length $L_x$ is deployed along the $x$-axis, with $M$ PAs mounted on the waveguide. The waveguide feed point is positioned at ${\mathbf r}_0=[0,0,d]^T$, where $d$ represents the height of the waveguide above the ground. Let the initial coordinate of the $m$-th PA be denoted as
\begin{equation}
	{\mathbf r}'_{m}=[x'_{m},0,d]^T, \  \forall m\in\mathcal M.
\end{equation}

\begin{figure}[!t]
	\centering
	\includegraphics[width=0.50\textwidth]{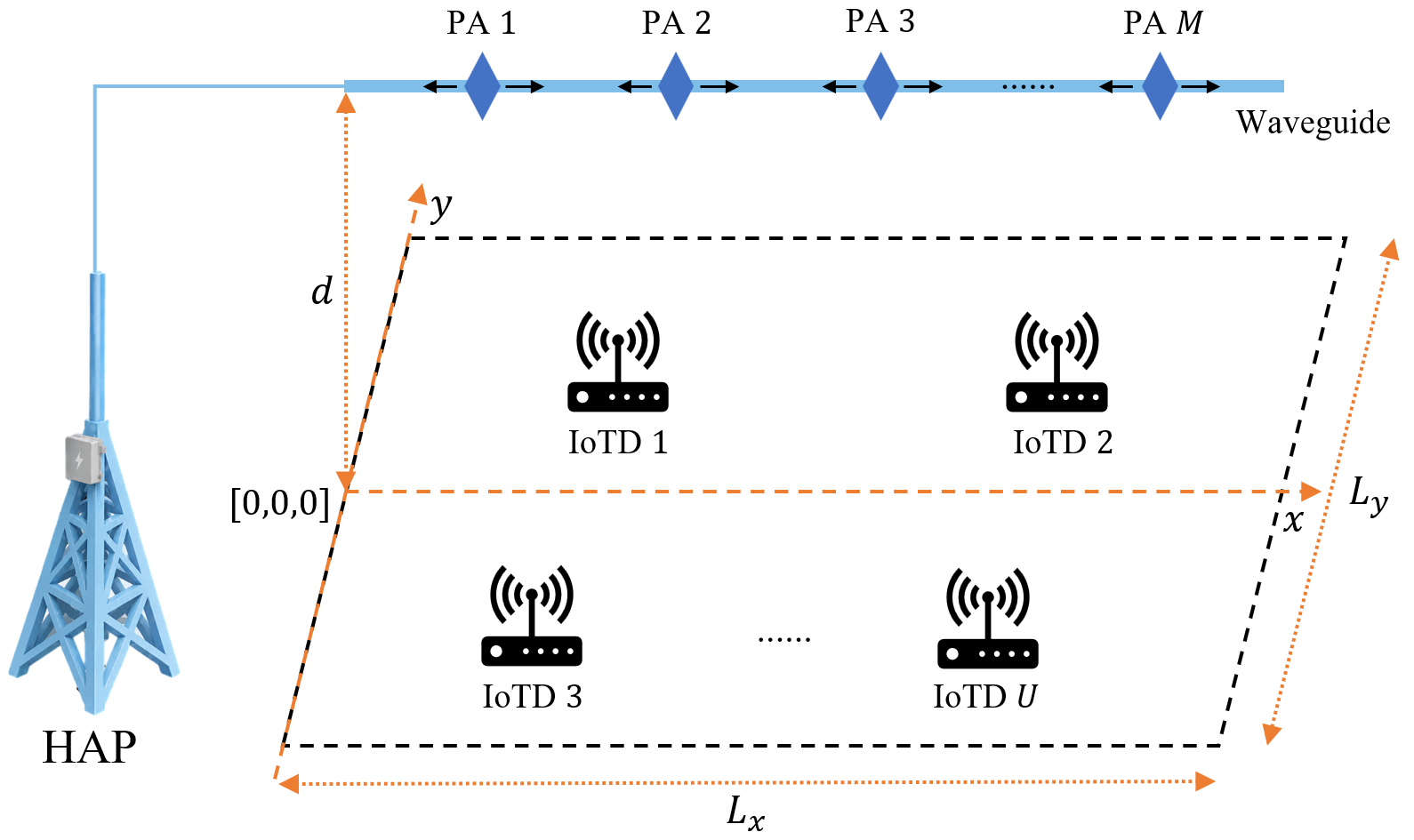}
	\caption{A typical PA-assisted WPCN system.}
	\label{Fig1}
	\vspace{-5pt}
\end{figure}

\begin{figure}[!t]
	\centering
	\includegraphics[width=0.45\textwidth]{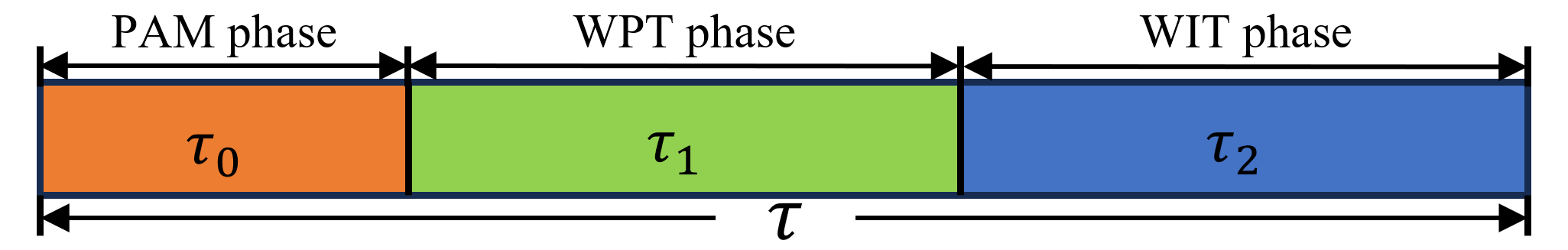}
	\caption{Illustration of the three-phase time block for the PA-assisted WPCN.}
	\label{Fig2}
	\vspace{-5pt}
\end{figure}

\vspace{-10pt}
\subsection{PAM Phase}
In the considered system, the PAs are allowed to dynamically adjust their positions along the waveguide during the PAM phase. By exploiting PA mobility, the WPCN performance can be improved. However, the PAM also incurs additional energy consumption and consumes part of the time block.

During the PAM phase, the $ m $-th PA is repositioned along the waveguide, and its coordinate after movement is expressed as
\begin{equation}
	{\mathbf r}_m= [x_m,0,d]^T, \ \forall m \in \mathcal M.
\end{equation}

The $U$ IoTDs are distributed on the ground within a rectangular region of dimensions $L_x\times L_y$, where the coordinate of IoTD $u$ is denoted as
\begin{equation}
	\bar{\mathbf r}_u= [\bar{x}_u,\bar{y}_u,0]^T, \  \forall u \in \mathcal U,
\end{equation}
where the IoTDs are assumed to remain stationary throughout the considered time block.

As the PAs are restricted to movement along the waveguide, only their $x$-coordinates vary during movement. Therefore, the movement distance of the $m$-th PA can be expressed as
\begin{equation}
	\Delta x_m=\left| x_m - x'_m \right|,\ \forall m\in\mathcal M.
\end{equation}
The movement of the PAs during the PAM duration $ \tau_0 $ is driven by electric motors, such as stepper motors \cite{M_Samy_2025}. Denote the speed of the PAs by $ v $, and the movement distance is constrained by
\begin{equation}
	\Delta x_m \le v\tau_0,\ \forall m\in\mathcal M.
\end{equation}
Denote the power consumption of the motor driving each PA by $P^{\rm PAM}_{\rm H}$. As a result, the total PAM energy consumption can be expressed as
\begin{equation}
	\Xi^{\rm PAM}_{\rm H} = P^{\rm PAM}_{\rm H} \sum\nolimits_{m=1}^{M} \frac{\Delta x_m}{v}.
\end{equation}

Moreover, to avoid mutual coupling between adjacent PAs and ensure reliable operation, the distance between any two neighboring PAs is required to be no smaller than a predefined threshold $\Delta_{\min}$, i.e.,
\begin{equation}
	x_{m+1}-x_m\ge \Delta_{\min},\ \forall m=1,\ldots,M-1.
\end{equation}
Furthermore, all PAs are constrained to move within the waveguide of length $L_x$, i.e., 
\begin{equation}
	0 \le x_m \le L_x,\  \forall m \in \mathcal{M}.
\end{equation}

\vspace{-15pt}
\subsection{WPT Phase}
During the WPT phase, with the designated positions of PAs, the HAP broadcasts RF energy to all IoTDs through the PAs. The near-field propagation is considered for the wireless channels between the PAs and the IoTDs due to the large antenna aperture of the PA-assisted system. Since the PA positions are fixed after the PAM phase and the IoTDs remain stationary, the downlink channels remain unchanged during both the WPT and WIT phases.

Denote the coordinates of feed point, the $m$-th PA, and IoTD $u$ by ${\mathbf r}_0$, ${\mathbf r}_m$, and $\bar{\mathbf r}_u$, respectively. Accordingly, the downlink channel between the $m$-th PA and IoTD $u$ is expressed as \cite{Y_Liu_2026_2}
\begin{equation}
	h_{m,u} =
	\frac{1}{\sqrt{M}}
	\frac{\lambda}{4\pi}
	\frac{
		e^{-j\frac{2\pi}{\lambda} d_{m,u}}
	}{
		d_{m,u}
	}
	e^{-j\frac{2\pi}{\lambda_g}
		d_{m}},
\end{equation}
where
\begin{equation}
	d_{m,u} = \|\mathbf r_m-\bar{\mathbf r}_u\| = \sqrt{(x_m-\bar{x}_u)^2+\bar{y}_u^2+d^2},
\end{equation}
and
\begin{equation}
	d_{m} = \|\mathbf r_m-\mathbf r_0\| = x_m.
\end{equation}
Here, $\lambda$ represents the free-space wavelength, while $\lambda_g=\lambda/n_{\rm eff}$ is the corresponding wavelength inside the waveguide. The term $1/\sqrt{M}$ indicates that the waveguide power is shared uniformly across the $M$ PAs\footnote{According to recent studies, PA systems commonly employ high-purity Teflon-based dielectric waveguides operating at 28 GHz, which exhibit a propagation loss of approximately 0.08 dB/m. This loss is negligible compared with the free-space propagation loss over the considered propagation distances \cite{C_Gao_2026, P_Liu_2026}. As a result, equal power allocation among the $M$ PAs can be adopted by controlling the power leakage of each PA \cite{Z_Wang_2025, Z_Ding_2025}.}.Accordingly, the equivalent downlink channel between the HAP and IoTD $u$ is given by
\begin{equation}
	h_u
	=
	\sum\nolimits_{m=1}^{M}
	h_{m,u},\  \forall u \in \mathcal{U}.
\end{equation}

Let $ p_{\rm H} $ denote the transmit power of the HAP, so the received RF power of IoTD $u$ is expressed as $ p_{\rm H} |h_u|^2 $. After receiving the RF signals, IoTD $ u $ converts them into direct current (DC) via its EH circuit. To capture the practical characteristics of RF-to-DC conversion, a nonlinear EH model is adopted \cite{E_Boshkovska_2015}. Accordingly, the harvested DC power at IoTD $ u $ is expressed by
\begin{equation}
	\label{deqn_ex1a}
	\Phi_u = \frac{\left[\frac{ \Phi_{\rm max} }{1 + {\rm exp}(-a(p_{\rm H} |h_u|^2-b)) } -  \Phi_{\rm max}\Omega\right]}{1 - \Omega},\  \forall u \in \mathcal{U},
\end{equation}
where $ \Phi_{\rm max} $ is a constant, which denotes the saturation output power of the EH circuit. The constant $ \Omega=\frac{1}{1+{\rm exp}(ab)} $ is introduced to ensure that zero input yields zero output, while $ a $ and $ b $ are fixed circuit-dependent parameters. Therefore, the energy harvested by IoTD $u$ within the WPT phase is $ \tau_1\Phi_u $.

As a result, the WPT energy consumption of the HAP is given by 
\begin{equation}
	\Xi_{\rm H}^{\rm WPT}
	= p_{\rm H}\tau_1.
\end{equation}

\subsection{WIT Phase}
During the WIT phase, all IoTDs utilize the harvested energy to transmit their information to the HAP. Since the uplink and downlink share the same waveguide, the reciprocal channel model is adopted. Accordingly, the equivalent uplink channel between IoTD $u$ and the HAP is identical to the corresponding downlink channel, which is expressed as
\begin{equation}
	g_u=h_u,\  \forall u \in \mathcal{U}.
\end{equation}
The signal received by the HAP is given by
\begin{equation}
	y
	=
	\sum\nolimits_{u=1}^{U}
	\sqrt{p_u}
	g_u
	s_u
	+
	n,
\end{equation}
where $n\sim\mathcal{CN}(0,\sigma^2)$ represents the additive white Gaussian noise (AWGN). In order to utilize limited spectrum efficiently, the uplink transmission employs the NOMA protocol, allowing multiple IoTDs to concurrently send their signals to the HAP. At the HAP, successive interference cancellation (SIC) is employed to separate the superimposed signals \cite{Q_Wu_2018}. Without loss of generality, the IoTDs are indexed in descending order of their normalized channel gains, i.e.,
\begin{equation}
	\gamma_U
	\ge
	\cdots
	\ge
	\gamma_1,
\end{equation}
where
\begin{equation}
	\gamma_u
	=
	\frac{|g_u|^2}{\sigma^2},\  \forall u \in \mathcal{U}.
\end{equation}
Accordingly, the achievable throughput (bits/Hz) of IoTD $u$ is given by
\begin{equation}
	R_u
	=
	\tau_2
	\log_2
	\left(
	1+
	\frac{p_u\gamma_u}
	{\sum\nolimits_{i=u+1}^{U}p_i\gamma_i+1}
	\right),\  \forall u \in \mathcal{U}.
\end{equation}
Accordingly, the total system throughput is given by
\begin{equation}
	\sum\nolimits_{u=1}^{U} R_u =
	\tau_2 \log_2\left( 1+ \sum\nolimits_{u=1}^{U} p_u\gamma_u \right).
\end{equation}

The WIT energy consumption of IoTD $u$ is given by
\begin{equation}
	\Xi^{\rm WIT}_u = p_u\tau_2,\  \forall u \in \mathcal{U}.
\end{equation}

\section{Problem Formulation}
In the considered PA-assisted WPCN system, PA mobility provides additional spatial degrees of freedom, which can improve the wireless channel conditions and consequently enhance the performance of the WPCN. However, moving the PAs incurs additional energy consumption and time overhead. This tradeoff motivates us to account for the costs of PAM when designing the PA-assisted WPCN. To this end, we propose a joint energy–time optimization for green PA-assisted WPCN design, where a total system energy consumption minimization problem is formulated by jointly optimizing the PA positions, PAM duration, WPT duration, WIT duration, and transmit powers of HAP and IoTDs while satisfying the QoS constraint. Let $\mathbf{x}=[x_1,\ldots,x_M]^T$ denote the PA position vector and $\mathbf{p}=[p_1,\ldots,p_U]^T$ denote the IoTD transmit power vector. The resulting optimization problem is formulated as follows
\begin{subequations}
	\begin{align}
		&(\mathbf{P_0}): \min_{\substack{\{
				\mathbf{x},\tau_0,\tau_1,\\ \tau_2,
				p_{\rm H},\mathbf{p}\}
		}}
		P^{\rm PAM}_{\rm H}
		\sum\nolimits_{m=1}^{M} \frac{\Delta x_m}{v}
		+ p_{\rm H}\tau_1 \\
		&\mathrm{s.t.}~\tau_2 \log_2 \left(1 + \sum\nolimits_{u=1}^{U} p_u \frac{|h_u|^2}{\sigma^2} \right) \ge U R_{\min}, \\
		&~~~~~p_u\tau_2 \le \tau_1\Phi_u, \ \forall u, \\
		&~~~~~\tau_0+\tau_1+\tau_2 \le \tau, \\
		&~~~~~0\le p_{\rm H}\le P_{\rm H}^{\max}, \\
		&~~~~~0\le p_u\le P_u^{\max},\ \forall u, \\
		&~~~~~\Delta x_m \le v\tau_0,\ \forall m, \\
		&~~~~~x_{m+1}-x_m \ge \Delta_{\min},\ \forall m=1,\ldots,M-1, \\
		&~~~~~0 \le x_m \le L_x,\ \forall m.
	\end{align}
\end{subequations}
In $(\mathbf{P_0})$, the objective function (22a) minimizes the HAP energy consumption, which consists of the energy consumed for PAM and WPT. Constraint (22b) guarantees that the average throughput of IoTDs is no less than the required minimum throughput $R_{\min}$ \cite{K_Chi_2019}. Constraint (22c) guarantees that the energy consumed for information transmission by each IoTD does not exceed its harvested RF energy. Constraint (22d) ensures that the durations of the PAM, WPT, and WIT phases do not exceed the considered time block. Constraints (22e) and (22f) specify the transmit power budgets of the HAP and IoTDs, respectively. Finally, constraints (22g)-(22i) impose the PAM, inter-PA distance, and waveguide boundary constraints, respectively.

\section{ADMM-based Algorithm}
The optimization problem $(\mathbf{P_0})$ jointly involves PA position optimization and WPCN resource allocation, which are strongly coupled through the channel gains. Specifically, changing the PA positions alters the channel gains, which are multiplicatively coupled with both the HAP transmit power and the IoTD transmit powers. Therefore, the PA position optimization and WPCN resource allocation need to be decoupled into two subproblems. However, the competition between the PAM and WPCN durations and the lack of incentive for moving PAs to reduce WPT energy consumption make the two subproblems difficult to optimize independently. To address these issues, we propose an ADMM-based algorithm, as illustrated in Fig. 3, which decouples the original problem into two subproblems corresponding to PA position optimization and WPCN resource allocation, respectively. Importantly, an auxiliary variable $\tilde{\tau}_0$ is introduced to duplicate the PAM duration $\tau_0$, and a consensus constraint $\tau_0=\tilde{\tau}_0$ is imposed to coordinate the two subproblems. Specifically, with $\{\tau_0, \mathbf{x}\}$ fixed, the WPCN resource allocation variables $\{\tilde{\tau}_0,\tau_1,\tau_2,p_{\rm H},\mathbf{p}\}$ are optimized via a two-dimensional search over $\tilde{\tau}_0$ and $\tau_1$. With the WPCN resource allocation variables fixed, $\{\tau_0, \mathbf{x}\}$ are optimized using a fine-tuning-based SCA algorithm. The two subproblems are then alternately optimized under the augmented Lagrangian framework until convergence.

\begin{figure*}[!t]
	\centering
	\includegraphics[width=1\textwidth]{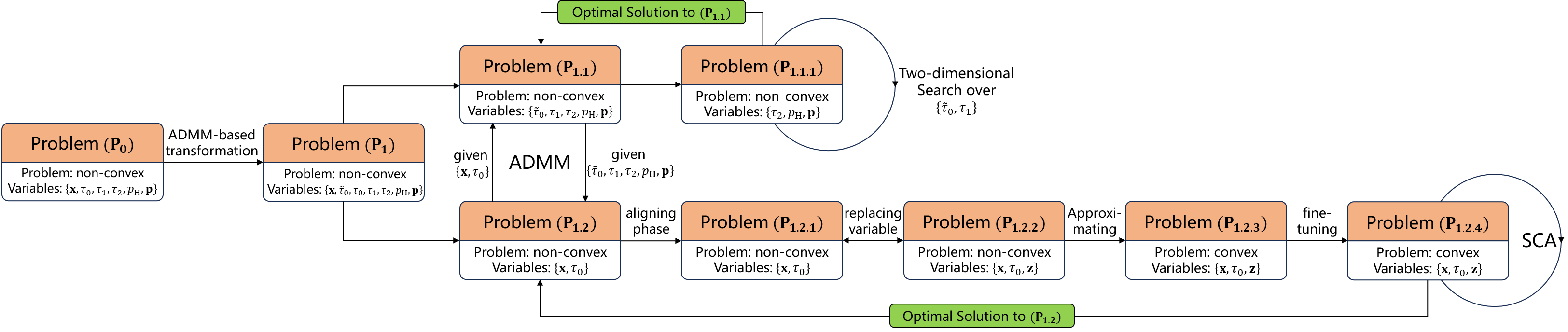}
	\caption{The flowchart of the ADMM-based algorithm.}
	\label{Fig3}
\end{figure*} 

\subsection{ADMM-based Transformation}
To facilitate the ADMM-based decomposition, an auxiliary variable $\tilde{\tau}_0=\tau_0$ is introduced as a copy of $\tau_0$, such that
\begin{equation} 
	\tilde{\tau}_0=\tau_0,\qquad 
	\tilde{\tau}_0+\tau_1+\tau_2\leq\tau.
\end{equation}
The constraint $\tilde{\tau}_0=\tau_0$ serves as the consensus constraint for the ADMM-based decomposition. Accordingly, $(\mathbf{P_0})$ is reformulated as
\begin{subequations}
	\begin{align}
		&(\mathbf{P_1}): \min_{\substack{\{\mathbf{x}, \tilde{\tau}_0,\tau_0, \tau_1,\\ \tau_2, p_{\rm H},\mathbf{p}\}}}
		P^{\rm PAM}_{\rm H}
		\sum\nolimits_{m=1}^{M} \frac{\Delta x_m}{v}
		+ p_{\rm H}\tau_1 \\
		&\mathrm{s.t.}~(22{\rm b}), (22{\rm c}), (22{\rm e}), (22{\rm f}), (22{\rm g}), (22{\rm h}), (22{\rm i}), \nonumber \\
		&~~~~~\tilde{\tau}_0+\tau_1+\tau_2 \le \tau, \\
		&~~~~~\tau_0 = \tilde{\tau}_0.
	\end{align}
\end{subequations}

To handle the consensus constraint $\tilde{\tau}_0=\tau_0$, we apply the ADMM framework \cite{Y_Wang_2019}. The corresponding augmented Lagrangian is given by
\begin{align}
	\mathcal{L}_{\rho}
	&\big(
	\mathbf{x},\tau_0,\tilde{\tau}_0,\tau_1,\tau_2,
	p_{\rm H},\mathbf{p},\nu
	\big) \nonumber \\
	&=P_{\rm H}^{\rm PAM}
	\sum_{m=1}^{M}\frac{\Delta x_m}{v}
	+p_{\rm H}\tau_1
	+
	\frac{\rho}{2}
	\left(
	\tau_0-\tilde{\tau}_0+\frac{\nu}{\rho}
	\right)^2,
\end{align}
where $\rho>0$ is the penalty parameter, and $\nu$ is the corresponding dual variable. The primal variables are updated alternately by minimizing the augmented Lagrangian over the two variable blocks, i.e., $\{\mathbf{x},\tau_0\}$ and $\{\tilde{\tau}_0,\tau_1,\tau_2,p_{\rm H},\mathbf{p}\}$. Given the updated primal variables at the $i$-th iteration, the dual variable is subsequently updated by
\begin{equation} 
	\nu^{(i)} = \nu^{(i-1)} + \rho \left(\tau_0^{(i)}-\tilde{\tau}_0^{(i)}\right). 
\end{equation}

So far, we have established the basic framework of the proposed ADMM-based algorithm. Next, the augmented Lagrangian $\mathcal{L}_{\rho}$ is minimized by alternately solving the PA position optimization subproblem over $\{\mathbf{x},\tau_0\}$ and the WPCN resource allocation subproblem with respect to $\{\tilde{\tau}_0,\tau_1,\tau_2,p_{\rm H},\mathbf{p}\}$.

\subsection{Resource Allocation to WPCN}
At the $i$-th iteration, given the PA position $\mathbf{x}^{(i-1)}$ and PAM duration $\tau_0^{(i-1)}$ obtained in the previous iteration, the movement distance of PA $m$ is denoted as $\Delta x_m(\mathbf{x}^{(i-1)})$, the channel from the HAP to IoTD $u$ is represented by $h_u(\mathbf{x}^{(i-1)})$, and the harvested DC power of IoTD $u$ is denoted as $\Phi_u(p_{\rm H},\mathbf{x}^{(i-1)})$, which depends on the HAP transmit power $p_{\rm H}$ and the channel power gain $|h_u(\mathbf{x}^{(i-1)})|^2$. Therefore, the WPCN resource allocation subproblem is expressed as
\begin{subequations}
	\begin{align}
		&(\mathbf{P_{1.1}}):
		\min_{\{\tilde{\tau}_0, \tau_1, \tau_2, p_{\rm H}, \mathbf{p}\}}
		P^{\rm PAM}_{\rm H}
		\sum\nolimits_{m=1}^{M} \frac{\Delta x_m(\mathbf{x}^{(i-1)})}{v}
		+ p_{\rm H}\tau_1 \nonumber \\ 
		&\qquad \qquad \qquad \qquad \quad+
		\frac{\rho}{2}
		\left(
		\tau_0^{(i-1)}-\tilde{\tau}_0+\frac{\nu^{(i-1)}}{\rho}
		\right)^2
		\\
		&\mathrm{s.t.}~\tau_2\log_2\left(1+\sum_{u=1}^{U}
		\frac{p_u|h_u(\mathbf{x}^{(i-1)})|^2}{\sigma^2}
		\right) \geq U R_{\min}, \\
		&~~~~~p_u\tau_2 \leq \tau_1\Phi_u(p_{\rm H}, \mathbf{x}^{(i-1)}),\ \forall u, \\
		&~~~~~(24{\rm b}), (22{\rm e}), (22{\rm f}). \nonumber
	\end{align}
\end{subequations}
With $\mathbf{x}^{(i-1)}$ given, the channel gains are fixed. Nevertheless, $(\mathbf{P_{1.1}})$ remains non-convex because of the variable coupling and the nonlinear EH model. Directly solving this problem using traditional non-convex optimization methods often involves approximations and iterations, which can lead to considerable computational complexity and make it difficult to guarantee a global optimum. 

To efficiently solve $(\mathbf{P_{1.1}})$, we reduce the original optimization problem to a two-dimensional search over the PAM duration copy $\tilde{\tau}_0$ and the WPT duration $\tau_1$. Specifically, we enumerate a candidate pair $(\tilde{\tau}_0,\tau_1)$ within its feasible range. Since the PA position vector $\mathbf{x}^{(i-1)}$ is fixed, the PAM energy consumption remains constant. Thus, $(\mathbf{P_{1.1}})$ reduces to minimizing the WPT energy consumption over the feasible range of $(\tilde{\tau}_0,\tau_1)$. For any given $(\tilde{\tau}_0,\tau_1)$, $(\mathbf{P_{1.1}})$ can be reformulated as
\begin{subequations}
	\begin{align}
		&(\mathbf{P_{1.1.1}}):
		\min_{\{\tau_2, p_{\rm H}, \mathbf{p}\}}
		p_{\rm H}\tau_1 
		\\
		&\mathrm{s.t.}~(27{\rm b}), (27{\rm c}), (24{\rm d}), (22{\rm e}), (22{\rm f}) \nonumber.
	\end{align}
\end{subequations}
By solving $ (\mathbf{P_{1.1.1}}) $, the corresponding optimal WIT duration $\tau_2^\star(\tilde{\tau}_0,\tau_1)$, the optimal HAP transmit power $p_{\rm H}^\star(\tilde{\tau}_0,\tau_1)$, and IoTD transmit powers ${p_u^\star(\tilde{\tau}_0,\tau_1)}$ can be obtained. The optimal pair $(\tilde{\tau}_0^\star,\tau_1^\star)$ is finally selected as the candidate that yields the minimum WPT energy consumption, i.e.,
\begin{align}
	(\tilde{\tau}_0^\star,\tau_1^\star)
	&=
	\arg\min_{\substack{\tilde{\tau}_0>0,\ \tau_1>0,\\
			\tilde{\tau}_0+\tau_1<\tau}}
	\Bigg[
	p_{\rm H}^\star(\tilde{\tau}_0,\tau_1)\tau_1
	\nonumber\\
	&\quad \quad \quad \quad \quad +
	\frac{\rho}{2}
	\left(
	\tau_0^{(i-1)}-\tilde{\tau}_0
	+\frac{\nu^{(i-1)}}{\rho}
	\right)^2
	\Bigg].
\end{align}

For a given pair $(\tilde{\tau}_0,\tau_1)$, the optimal WPCN resource allocation can be characterized by the following lemma.

\newtheorem{lemma}{Lemma}
\begin{lemma}
For a given pair $(\tilde{\tau}_0,\tau_1)$, its corresponding optimal WIT duration, IoTD transmit power, and HAP transmit power are respectively given by
\begin{equation}
	\tau_2^\star(\tilde{\tau}_0,\tau_1)=\tau-\tilde{\tau}_0-\tau_1,
\end{equation}
\begin{equation}
	p_u^\star(\tilde{\tau}_0,\tau_1)
	=
	\min\left\{
	P_u^{\max},
	\frac{\tau_1}{\tau_2}
	\Phi_u\big(p_{\rm H}^\star(\tilde{\tau}_0,\tau_1),\mathbf{x}^{(i-1)}\big)
	\right\},\ \forall u,
\end{equation}
and
\begin{equation}
	p_{\rm H}^\star(\tilde{\tau}_0,\tau_1)
	=
	\min\left\{
	p_{\rm H}:
	R(p_{\rm H})\geq U R_{\min},
	\ 0\leq p_{\rm H}\leq P_{\rm H}^{\max}
	\right\},
\end{equation}
where $R(p_{\rm H})$ denotes the system throughput.
\end{lemma}

\begin{IEEEproof}
First, for a fixed $(\tilde{\tau}_0,\tau_1)$, increasing $\tau_2$ allows the required QoS to be achieved with lower IoTD transmission rates and, consequently, lower IoTD transmit powers. This reduces the harvested energy required from the HAP and hence the required HAP transmit power $p_{\rm H}$. Therefore, the optimal WIT duration should occupy all the remaining time, yielding (30).

For a given pair $(\tau_1,\tau_2)$, the harvested energy of each IoTD depends only on the HAP transmit power $p_{\rm H}$ through the EH model. Since the achievable throughput is monotonically increasing in each $p_u$, each IoTD should transmit at the largest power permitted by its power budget and available harvested energy, i.e., (31).

Finally, the harvested power $\Phi_u(p_{\rm H},\mathbf{x}^{(i-1)})$ is monotonically non-decreasing with respect to $p_{\rm H}$. Therefore, the corresponding optimal $p_u$ is also monotonically non-decreasing with respect to $p_{\rm H}$, and hence the system throughput $R(p_{\rm H})$ is monotonically non-decreasing with $p_{\rm H}$. Since $\tau_1$ is fixed, reducing $p_{\rm H}$ directly lowers the objective $p_{\rm H}\tau_1$. Hence, the optimal $p_{\rm H}$ is obtained as the smallest transmit power satisfying the QoS constraint, i.e., (32). It can be efficiently determined by searching over $[0,P_{\rm H}^{\max}]$ with the bisection method, where the QoS constraint is evaluated according to $R(p_{\rm H})$.
\end{IEEEproof}

\subsection{PA Position Optimization}
Given the WPCN resource allocation
$\{\tilde{\tau}_0^{(i)},\tau_1^{(i)},\tau_2^{(i)},p_{\rm H}^{(i)},\mathbf{p}^{(i)}\}$
obtained from $(\mathbf{P_{1.1}})$, the PA position optimization subproblem is cast as
\begin{subequations}
	\begin{align}
		&(\mathbf{P_{1.2}}):
		\min_{\{\mathbf{x},\tau_0, p_{\rm H}\}}
		P_{\rm H}^{\rm PAM}
		\sum\nolimits_{m=1}^{M}
		\frac{\Delta x_m}{v}+p_{\rm H}\tau_1^{(i)}
		\nonumber \\
		&\qquad \qquad \qquad \qquad \qquad +
		\frac{\rho}{2}
		\left(
		\tau_0-\tilde{\tau}_0^{(i)}+\frac{\nu^{(i-1)}}{\rho}
		\right)^2
		\\
		&\mathrm{s.t.}~\tau_2^{(i)} \log_2\left(1+\sum\nolimits_{u=1}^{U}\frac{p_u^{(i)}|h_u|^2}{\sigma^2}
		\right)\geq U R_{\min}, \\
		&~~~~~p_u^{(i)}\tau_2^{(i)}\leq\tau_1^{(i)}\Phi_u,\ \forall u, \\
		&~~~~~(22{\rm e}), (22{\rm g}), (22{\rm h}), (22{\rm i}) \nonumber.
	\end{align}
\end{subequations}
Although the WPCN resource allocation variables are fixed in $(\mathbf{P_{1.2}})$, the HAP transmit power $p_{\rm H}$ is retained as an optimization variable. This treatment is necessary to provide an effective incentive for PAM. Specifically, if $p_{\rm H}$ were fixed at its value obtained from the WPCN resource allocation subproblem, the current PA positions may already satisfy the constraints. Since PAM incurs additional energy consumption, the position optimization would have no incentive to reposition the PAs, and the optimal solution would tend to remain at their initial positions. By allowing $p_{\rm H}$ to be jointly optimized with the PA positions, the PA position optimization can exploit the energy tradeoff between PAM and WPT\footnote{It should be emphasized that $p_{\rm H}$ in $(\mathbf{P_{1.2}})$ serves only as an auxiliary optimization variable to guide the PA position optimization and is not adopted as the final WPT transmit power.}. 

Problem $(\mathbf{P_{1.2}})$ is non-convex, as the channel power gain $|h_u|^2$ exhibits neither convexity nor concavity with respect to the PA positions. Specifically, PA positions simultaneously alter large-scale path loss and small-scale phase shift. Therefore, directly optimizing the PA positions requires jointly handling the distance-dependent path loss and the highly position-sensitive phase shift, which makes the resulting optimization difficult to solve.

To obtain a tractable solution, we propose a distance-phase decoupling strategy, i.e., a fine-tuning-based SCA algorithm, which separately considers the effects of the path loss and phase shift. Specifically, we begin by assuming perfect phase alignment among the signals radiated by different PAs and focus on the distance-dependent path-loss effect. Under this assumption, the original PA position optimization problem is simplified into a more tractable yet still non-convex problem, and subsequently, the SCA technique is used to obtain a tractable solution \cite{X_Zhang_2025}. After obtaining the coarse PA positions, we further consider the small-scale phase effect and refine the PA positions through a fine-tuning procedure, such that the phase alignment among the signals from different PAs can be maintained as much as possible \cite{C_Gao_2026}.

\subsubsection{Coarse PA position optimization focusing on path loss}
In the first stage, we focus on the distance-dependent path-loss effect while temporarily disregarding the impact of the position-dependent signal phase. Specifically, according to the triangle inequality, the channel power gain can be upper bounded as
\begin{align}
	|h_u|^2
	&=
	\frac{1}{M}
	\left(\frac{\lambda}{4\pi}\right)^2
	\left|
	\sum\nolimits_{m\in\mathcal{M}}
	\frac{1}{d_{m,u}}
	e^{-j\frac{2\pi}{\lambda}d_{m,u}}
	e^{-j\frac{2\pi}{\lambda_g}d_m}
	\right|^2
	\nonumber\\
	&\leq
	\frac{1}{M}
	\left(\frac{\lambda}{4\pi}\right)^2
	\left(\sum\nolimits_{m\in\mathcal{M}}
	\frac{1}{d_{m,u}}\right)^2.
\end{align}
The equality holds when the signals emitted by all PAs arrive at IoTD $ u $ with the same phase. Therefore, by temporarily assuming perfect phase alignment, we focus on the distance-dependent path-loss effect and adopt the following simplified channel power gain:
\begin{equation}
	\widetilde{|h_u|^2}
	=
	\frac{1}{M}
	\left(\frac{\lambda}{4\pi}\right)^2
	\left(\sum\nolimits_{m\in\mathcal{M}}
	\frac{1}{d_{m,u}}\right)^2.
\end{equation}
Accordingly, by replacing $ |h_u|^2 $ with
$\widetilde{|h_u|^2}$ and transforming constraints (33b) and (33c) into distance-dependent forms, the PA position optimization problem is reformulated as
\begin{subequations}
	\begin{align}
		&(\mathbf{P_{1.2.1}}):~
		\min_{\{\mathbf{x}, \tau_0, p_{\rm H}\}}
		P_{\rm H}^{\rm PAM}
		\sum\nolimits_{m=1}^{M}
		\frac{\Delta x_m}{v}
		+p_{\rm H}\tau_1^{(i)}
		\nonumber \\
		&\qquad \qquad \qquad \qquad \qquad +
		\frac{\rho}{2}
		\left(
		\tau_0-\tilde{\tau}_0^{(i)}+\frac{\nu^{(i-1)}}{\rho}
		\right)^2
		\\
		&\mathrm{s.t.}~\sum\nolimits_{u=1}^{U} p_u^{(i)} \widetilde{|h_u|^2} \geq \sigma^2 \left( 2^{\frac{U R_{\min}}{\tau_2^{(i)}}}-1\right), \\ 
		&~~~~~\widetilde{|h_u|^2} \geq \frac{\Gamma_u^{(i)}}{p_{\rm H}},\ \forall u, \\
		&~~~~~(22{\rm e}), (22{\rm g}), (22{\rm h}), (22{\rm i}). \nonumber
	\end{align}
\end{subequations} 
where $ \Gamma_u^{(i)} $ represents the minimum received power threshold required by IoTD $ u $ to satisfy its energy harvesting requirement and is expressed as
\begin{equation}
	\Gamma_u^{(i)}
	\triangleq
	b+
	\frac{1}{a}
	\ln
	\left(
	\frac{
		\Omega+
		\frac{(1-\Omega)p_u^{(i)}\tau_2^{(i)}}
		{\Phi_{\max}\tau_1^{(i)}}
	}{
		1-\Omega-
		\frac{(1-\Omega)p_u^{(i)}\tau_2^{(i)}}
		{\Phi_{\max}\tau_1^{(i)}}
	}
	\right).
\end{equation}

Although the phase effect is temporarily neglected, $(\mathbf{P_{1.2.1}})$ is still non-convex because $\widetilde{|h_u|^2}$ contains the inverse-distance terms $1/d_{m,u}$, which are non-convex functions of the PA positions. Therefore, a further SCA method is required to obtain a tractable solution to $(\mathbf{P_{1.2.1}})$.

To facilitate the subsequent SCA, an auxiliary variable $z_{m,u}$ is employed to decouple the inverse-distance term from the PA positions. Let $ \mathbf{z} \triangleq \left\{z_{m,u}\mid m\in\mathcal{M},u\in\mathcal{U}\right\} $, based on Proposition 1, $(\mathbf{P_{1.2.1}})$ is equivalently transformed into
\begin{subequations}
	\begin{align}
		&(\mathbf{P_{1.2.2}}):~
		\min_{\{\mathbf{x}, \tau_0, \mathbf{z},p_{\rm H}\}}
		P_{\rm H}^{\rm PAM}
		\sum\nolimits_{m=1}^{M}
		\frac{\Delta x_m}{v}
		+p_{\rm H}\tau_1^{(i)}
		\nonumber \\
		&\qquad \qquad \qquad \qquad \qquad +
		\frac{\rho}{2}
		\left(
		\tau_0-\tilde{\tau}_0^{(i)}+\frac{\nu^{(i-1)}}{\rho}
		\right)^2
		\\
		&\mathrm{s.t.}~\sum_{u=1}^{U}
		\frac{p_u^{(i)}}{M} 
		\left(\frac{\lambda}{4\pi}\right)^2
		\left(\sum_{m\in\mathcal{M}}z_{m,u}^{-\frac{1}{2}}\right)^2
		\geq
		\sigma^2
		\left(2^{\frac{U R_{\min}}{\tau_2^{(i)}}}-1\right), \\
		&~~~~~
		\frac{1}{M}
		\left(\frac{\lambda}{4\pi}\right)^2
		\left(\sum\nolimits_{m\in\mathcal{M}}z_{m,u}^{-\frac{1}{2}}\right)^2
		\geq
		\frac{\Gamma_u^{(i)}}{p_{\rm H}},
		\ \forall u, \\
		&~~~~~(x_m-\bar{x}_u)^2+\bar{y}_u^2+d^2 \leq z_{m,u},
		\ \forall m,u, \\
		&~~~~~(22{\rm e}), (22{\rm g}), (22{\rm h}), (22{\rm i}). \nonumber
	\end{align}
\end{subequations}

\newtheorem{proposition}{Proposition}
\begin{proposition}
	Problems $(\mathbf{P_{1.2.1}})$ and $(\mathbf{P_{1.2.2}})$ are equivalent.
\end{proposition}

\begin{IEEEproof}
	By imposing
	\begin{equation}
		d_{m,u}^{2 }= (x_m-\bar{x}_u)^2+\bar{y}_u^2+d^2 \leq z_{m,u},
	\end{equation}
	we have
	\begin{equation}
		\frac{1}{d_{m,u}} \geq z_{m,u}^{-\frac{1}{2}}.
	\end{equation}
	Accordingly, we obtain the following lower-bound expression for $ \widetilde{|h_u|^2} $:
	\begin{equation}
		\widetilde{|h_u|^2}
		\geq
		\frac{1}{M}
		\left(\frac{\lambda}{4\pi}\right)^2
		\left(\sum\nolimits_{m\in\mathcal{M}}z_{m,u}^{-\frac{1}{2}}\right)^2.
	\end{equation}
	
	Therefore, any feasible solution to $(\mathbf{P_{1.2.2}})$ is also feasible for $(\mathbf{P_{1.2.1}})$ when projected onto the original variables. Conversely, for any feasible solution to $(\mathbf{P_{1.2.1}})$, setting $z_{m,u}=d_{m,u}^2$ makes the above inequality hold with equality, ensuring the feasibility of $(\mathbf{P_{1.2.2}})$. Since the objective functions of the two problems are identical and do not depend on $\mathbf{z}$, the two problems have the same optimal objective value. This completes the proof.
\end{IEEEproof}

Although the auxiliary variables $ z_{m,u} $ provide a convenient representation of the inverse-distance term, $ (\mathbf{P_{1.2.2}}) $ is still non-convex. Specifically, $\left(\sum\nolimits_{m\in\mathcal{M}}z_{m,u}^{-\frac{1}{2}}\right)^2$ is a convex function of $ z_{m,u} $, while the corresponding constraints involve their superlevel sets. Therefore, these constraints are non-convex, and an additional SCA procedure is required to obtain a tractable convex approximation of $ (\mathbf{P_{1.2.2}}) $.

The non-convexity of $(\mathbf{P_{1.2.2}})$ is handled using the SCA framework. Specifically, at the $ k $-th SCA iteration, we construct a first-order Taylor approximation of the convex function $\left(\sum\nolimits_{m\in\mathcal{M}}z_{m,u}^{-\frac{1}{2}}\right)^2$ around the local point $z_{m,u}^{(k-1)}$. Specifically, we have
\begin{align}
	\widetilde{|h_u|^2}
	&\geq
	\widetilde{|h_u|^2}_{\mathrm{lb}}^{(k-1)},
\end{align}
where
\begin{align}
	&\widetilde{|h_u|^2}_{\mathrm{lb}}^{(k-1)}
	=
	\frac{1}{M}
	\left(\frac{\lambda}{4\pi}\right)^2
	\Bigg[
	\left(
	\sum\nolimits_{m\in\mathcal{M}}
	(z_{m,u}^{(k-1)})^{-\frac12}
	\right)^2
	\nonumber\\
	&-
	\left(
	\sum_{m\in\mathcal{M}}
	(z_{m,u}^{(k-1)})^{-\frac12}
	\right)
	\sum_{m\in\mathcal{M}}
	(z_{m,u}^{(k-1)})^{-\frac32}
	\left(
	z_{m,u}-z_{m,u}^{(k-1)}
	\right)
	\Bigg].
\end{align}
Accordingly, by replacing $\widetilde{|h_u|^2}$ with its first-order Taylor lower bound $ \widetilde{|h_u|^2}_{\mathrm{lb}}^{(k-1)} $, $(\mathbf{P_{1.2.2}})$ can be approximated by
\begin{subequations}
	\begin{align}
		&(\mathbf{P_{1.2.3}}):~
		\min_{\{\mathbf{x},\tau_0,\mathbf{z},p_{\rm H}\}}
		P_{\rm H}^{\rm PAM}
		\sum\nolimits_{m=1}^{M}
		\frac{\Delta x_m}{v}
		+p_{\rm H}\tau_1^{(i)}
		\nonumber \\
		&\qquad \qquad \qquad \qquad \qquad +
		\frac{\rho}{2}
		\left(
		\tau_0-\tilde{\tau}_0^{(i)}+\frac{\nu^{(i-1)}}{\rho}
		\right)^2
		\\
		&\mathrm{s.t.}~\sum\nolimits_{u=1}^{U}
		p_u^{(i)} \widetilde{|h_u|^2}_{\mathrm{lb}}^{(k-1)} \geq
		\sigma^2
		\left(
		2^{\frac{U R_{\min}}{\tau_2^{(i)}}}-1
		\right),
		\\
		&~~~~~
		\widetilde{|h_u|^2}_{\mathrm{lb}}^{(k-1)} \geq
		\frac{\Gamma_u^{(i)}}{p_{\rm H}},
		\ \forall u,
		\\
		&~~~~~
		z_{m,u} \geq
		(x_m-\bar{x}_u)^2+\bar{y}_u^2+d^2,
		\ \forall m,u. \\
		&~~~~~(22{\rm e}), (22{\rm g}), (22{\rm h}), (22{\rm i}). \nonumber
	\end{align}
\end{subequations}
Since the first-order Taylor approximation of $\widetilde{|h_u|^2}$ is the affine function of $z_{m,u}$, constraints (44b), (44c), and (44d) are convex. Therefore, $(\mathbf{P_{1.2.3}})$ is a convex optimization problem.

\subsubsection{Fine PA position adjustment focusing on phase shift}
After obtaining the coarse PA positions by solving $(\mathbf{P_{1.2.3}})$, we further consider the position-dependent phase shift to refine the PA positions. To achieve coherent combining of the signals arriving at IoTD $u$ from different PAs, the phase-alignment condition can be expressed as
\begin{equation}
	d_{m,u}+n_{\rm eff}d_m=p_{m,u}\lambda,
	\ \forall m,u,
\end{equation}
where $p_{m,u}$ is a positive integer.

In general, the above phase-alignment condition cannot be fulfilled for every PA and IoTD at the same time, because the required PA positions vary across different IoTDs. We therefore refine the coarse PA positions by minimizing the phase mismatch between the total propagation path and its nearest integer multiple of the wavelength. Specifically, define
\begin{equation}
	f_{m,u}(x_m)
	\triangleq
	\sqrt{(x_m-\bar{x}_u)^2+\bar{y}_u^2+d^2}
	+n_{\rm eff}x_m.
\end{equation}
Accordingly, the phase mismatch can be characterized by
\begin{equation}
	\sum\nolimits_{m\in\mathcal M}
	\sum\nolimits_{u\in\mathcal U}
	\left|
	f_{m,u}(x_m)-p_{m,u}\lambda
	\right|.
\end{equation}
The integer-valued nature of $p_{m,u}$ makes the above objective non-convex, preventing its direct optimization. To obtain a tractable
form, we determine the integer $p_{m,u}$ according to the PA position
obtained in the previous iteration and approximate $f_{m,u}(x_m)$ by
its first-order Taylor expansion at $x_m^{(k-1)}$, i.e.,
\begin{equation}
	f_{m,u}(x_m)
	\approx
	f_{m,u}(x_m^{(k-1)})
	+
	f_{m,u}'(x_m^{(k-1)})
	\left(x_m-x_m^{(k-1)}\right),
\end{equation}
where
\begin{equation}
	f_{m,u}'(x_m^{(k-1)})
	=
	\frac{x_m^{(k-1)}-\bar{x}_u}
	{\sqrt{(x_m^{(k-1)}-\bar{x}_u)^2+\bar{y}_u^2+d^2}}
	+n_{\rm eff}.
\end{equation}
Accordingly, the phase-mismatch penalty is approximated as
\begin{align}
	&g_{m,u}^{(k-1)}(x_m)
	=
	\Bigg|
	f_{m,u}(x_m^{(k-1)})
	+
	f_{m,u}'(x_m^{(k-1)})
	\left(x_m-x_m^{(k-1)}\right)
	\nonumber\\
	&\qquad \qquad \qquad \qquad \qquad \qquad-
	\left\lfloor
	\frac{f_{m,u}(x_m^{(k-1)})}{\lambda}
	\right\rceil \lambda
	\Bigg|.
\end{align}

To account for the phase-alignment performance while preserving the PA position design for path-loss reduction, we incorporate the phase-mismatch penalty into the objective function of $(\mathbf{P_{1.2.3}})$. Specifically, with a weighting factor $\eta\geq0$, the refined PA position optimization problem is formulated as
\begin{subequations}
	\begin{align}
		&(\mathbf{P_{1.2.4}}):~
		\min_{\{\mathbf{x},\tau_0,\mathbf{z},p_{\rm H}\}}
		P_{\rm H}^{\rm PAM}
		\sum\nolimits_{m=1}^{M}
		\frac{\Delta x_m}{v}
		+p_{\rm H}\tau_1^{(i)} 
		\nonumber \\
		& +
		\frac{\rho}{2}
		\left(
		\tau_0-\tilde{\tau}_0^{(i)}+\frac{\nu^{(i-1)}}{\rho}
		\right)^2 + \eta
		\sum_{m\in\mathcal M}
		\sum_{u\in\mathcal U}
		g_{m,u}^{(k-1)}(x_m)
		\\
		&\mathrm{s.t.}~(44{\rm b}), (44{\rm c}), (44{\rm d}), (22{\rm e}), (22{\rm g}), (22{\rm h}), (22{\rm i}). \nonumber
	\end{align}
\end{subequations}
Since $g_{m,u}^{(k-1)}(x_m)$ is the absolute value of an affine function of $x_m$, it is convex. Consequently, $(\mathbf{P_{1.2.4}})$ is convex and admits an efficient solution via standard convex optimization tools. The obtained solution is then used to update the local point for the next SCA iteration until convergence. 

For clarity, the proposed ADMM-based algorithm is summarized in Algorithm 1. In each ADMM iteration, the WPCN resource allocation block and the PA position optimization block are alternately updated to enforce the consensus constraint $\tau_0=\tilde{\tau}_0$. After convergence, the obtained PA positions are substituted into the original channel model, based on which the WPCN resource allocation is re-optimized to obtain the final system energy consumption.

\begin{algorithm}[t]
	\caption{ADMM-Based Algorithm}
	\begin{algorithmic}[1]
		\State \textbf{Initialize:} $\mathbf{x}^{(0)}$, $\tau_0^{(0)}$, $\nu^{(0)}$, $\rho$, $\kappa$, and $i\gets1$;
		
		\Repeat
		
		\State \textbf{\%WPCN Resource Allocation:}
		\For{each $\tilde{\tau}_0\in(0,\tau)$}
		\For{each $\tau_1\in(0,\tau-\tilde{\tau}_0)$}
		\State Obtain $\tau_2^\star(\tilde{\tau}_0,\tau_1)$ according to (30);
		\State Obtain $p_{\rm H}^\star(\tilde{\tau}_0,\tau_1)$ according to (32);
		\State Obtain $p_u^\star(\tilde{\tau}_0,\tau_1)$ according to (31), $\forall u$;
		\EndFor
		\EndFor
		
		\State Obtain $(\tilde{\tau}_0^\star,\tau_1^\star)$ according to (29);
		
		\State $(\tilde{\tau}_0^{(i)},\tau_1^{(i)}) \gets (\tilde{\tau}_0^\star,\tau_1^\star)$;
		\State $\tau_2^{(i)} \gets \tau_2^\star(\tilde{\tau}_0^\star,\tau_1^\star)$;
		\State $p_{\rm H}^{(i)} \gets p_{\rm H}^\star(\tilde{\tau}_0^\star,\tau_1^\star)$,
		$p_u^{(i)} \gets p_u^\star(\tilde{\tau}_0^\star,\tau_1^\star)$;

		\State \textbf{\%PA Position Optimization:}
		
		\State Given $\mathbf{x}_{\rm SCA}^{(0)}\gets\mathbf{x}^{(i-1)}$ and
		$\mathbf{z}_{\rm SCA}^{(0)}$ as an initial solution;
		
		\State Initialize SCA iteration index $k\gets1$;
		
		\Repeat
		
		\State Obtain $(\mathbf{x}^\star, \tau_0^\star, \mathbf{z}^\star)$ by solving $(\mathbf{P_{1.2.4}})$;
		
		\State Update $\mathbf{x}_{\rm SCA}^{(k)} \gets \mathbf{x}^\star$;
		
		\State Update $\mathbf{z}_{\rm SCA}^{(k)} \gets \mathbf{z}^\star$;
		
		\State Update $k\gets k+1$;
		
		\Until{The preset SCA stopping condition is reached}
		
		\State $(\mathbf{x}^{(i)},\tau_0^{(i)}) \gets (\mathbf{x}^\star,\tau_0^\star)$;

		\State Update $ \nu^{(i)} = \nu^{(i-1)} + \rho \left(\tau_0^{(i)}-\tilde{\tau}_0^{(i)}\right) $;
		
		\State Update $ \rho \gets \kappa \rho $;
		
		\State $i\gets i+1$;
		
		\Until{The preset ADMM stopping condition is reached}
		
		\State \Return
		$\left\{
		\mathbf{x}^{\star},
		\tau_0^{\star},
		\tau_1^{\star},
		\tau_2^{\star},
		p_{\rm H}^{\star},
		\mathbf{p}^{\star}
		\right\}$
	\end{algorithmic}
\end{algorithm}

\vspace{-10pt}
\subsection{Convergence and Computational Complexity}
\subsubsection{Convergence}
The convergence of the proposed ADMM-based algorithm is analyzed through its two subproblems. Since $(\mathbf{P_{1.2.4}})$ is a convex approximation of the original minimization problem $(\mathbf{P_{1.2}})$, its objective value provides an upper bound that is monotonically tightened over SCA iterations for a fixed penalty parameter. Meanwhile, $(\mathbf{P_{1.1}})$ is optimally solved at each iteration. By alternating between the two subproblems, the augmented Lagrangian function is non-increasing and bounded from below, ensuring convergence to a stationary point of the original problem.

\subsubsection{Computational Complexity}
The computational complexity of the proposed ADMM-based algorithm is determined by the computational costs of solving the two subproblems, i.e., WPCN resource allocation and PA position optimization. The former is solved by a two-dimensional search over $\tilde{\tau}_0$ and $\tau_1$ with a computational complexity of $\mathcal{O}(N_{\tilde{\tau}_0}N_{\tau_1})$, where $N_{\tilde{\tau}_0}$ and $N_{\tau_1}$ denote the numbers of candidate values of $\tilde{\tau}_0$ and $\tau_1$, respectively. The latter is solved using the SCA method, where each iteration employs an interior-point method to solve the resulting convex problem \cite{K_-Y_Wang_2014}. Since the convex problem involves $(MU+M+2)$ variables, its computational complexity is $\mathcal{O}((MU+M+2)^3)$ per SCA iteration. Let $N_{\rm SCA}$ denote the number of SCA iterations. Thus, the computational cost of this subproblem is about $\mathcal{O}(N_{\rm SCA}(MU+M+2)^3)$. Let $I_{\rm ADMM}$ denote the number of ADMM iterations. Hence, the total computational complexity is about $ \mathcal{O}\!\left[I_{\rm ADMM}\left( N_{\tilde{\tau}_0}N_{\tau_1}+N_{\rm  SCA}(MU+M+2)^3\right)\right] $.

\section{AO-based Algorithm for Special Case}
In this section, we consider a special case where the PAM duration is predetermined by the motor operation protocol, hardware control cycle, or system-level scheduling requirements \cite{Y_Xu_2026_2}. Under such a fixed PAM duration, the key issue is balancing the energy tradeoff between PAM and WPCN. The resulting optimization problem is formulated as follows
\begin{subequations}
	\begin{align}
		&(\mathbf{P_2}): \min_{\substack{\{
				\mathbf{x},\tau_1,\\ \tau_2,
				p_{\rm H},\mathbf{p}\}
		}}
		P^{\rm PAM}_{\rm H}
		\sum\nolimits_{m=1}^{M} \frac{\Delta x_m}{v}
		+ p_{\rm H}\tau_1 \\
		&\mathrm{s.t.}~(22{\rm b}), (22{\rm c}), (22{\rm d}), (22{\rm e}), (22{\rm f}), (22{\rm g}), (22{\rm h}), (22{\rm i}) \nonumber.
	\end{align}
\end{subequations}
Different from the general case, the PAM duration $\tau_0$ is fixed in $(\mathbf{P_2})$. Therefore, the coupling caused by the shared PAM duration between PA position optimization and WPCN resource allocation is eliminated, and no auxiliary variable or consensus constraint is required. Consequently, an AO-based algorithm with lower computational complexity is proposed. Specifically, $(\mathbf{P_2})$ is decomposed into two subproblems, i.e., PA position optimization and WPCN resource allocation, which are alternately optimized while fixing the variables of the other subproblem. The overall workflow of the proposed AO-based algorithm is illustrated in Fig. 4.

\begin{figure}[!t]
	\centering
	\includegraphics[width=0.4\textwidth]{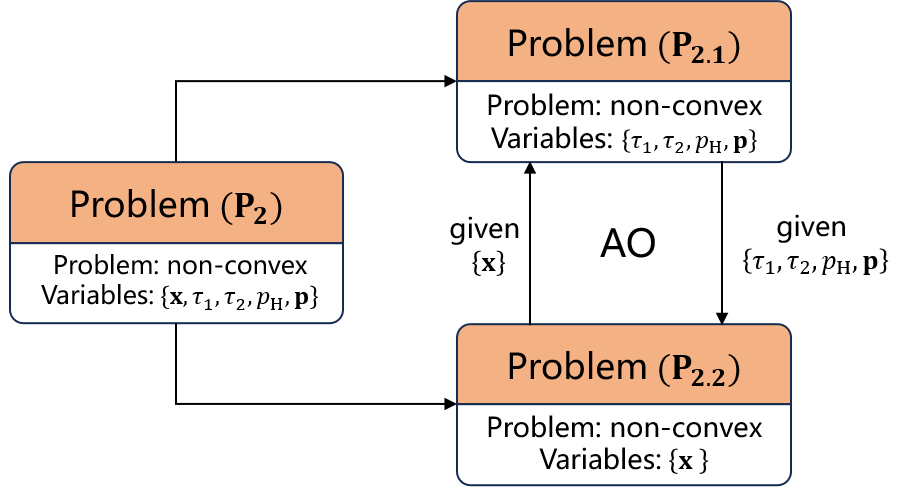}
	\caption{The flowchart of the AO-based algorithm.}
	\label{Fig4}
	\vspace{-15pt}
\end{figure} 

\subsection{Resource Allocation to WPCN for Special Case}
At the $i$-th iteration, given the updated PA positions $\mathbf{x}^{(i-1)}$, the WPCN resource allocation block $\{\tau_1^{(i)},\tau_2^{(i)},p_{\rm H}^{(i)},\mathbf{p}^{(i)}\}$ is updated by solving
\begin{subequations}
	\begin{align}
		&(\mathbf{P_{2.1}}):
		\min_{\{\tau_1, \tau_2, p_{\rm H}, \mathbf{p}\}}
		p_{\rm H}\tau_1
		\\
		&\mathrm{s.t.}~\tau_2\log_2\left(1+\sum_{u=1}^{U}
		\frac{p_u|h_u(\mathbf{x}^{(i-1)})|^2}{\sigma^2}
		\right) \geq U R_{\min}, \\
		&~~~~~p_u\tau_2 \leq \tau_1\Phi_u(p_{\rm H}, \mathbf{x}^{(i-1)}),\ \forall u, \\
		&~~~~~(22{\rm d}), (22{\rm e}), (22{\rm f}). \nonumber
	\end{align}
\end{subequations}
Since the PAM duration $\tau_0$ is fixed, the WPCN resource allocation problem does not involve the auxiliary variable or the ADMM penalty term. Therefore, $(\mathbf{P_{2.1}})$ can be efficiently solved through a one-dimensional search over $\tau_1$. For each candidate $\tau_1$, the optimal resource allocation is obtained as in Section~IV-A, and $\tau_1^\star$ is selected to minimize $p_{\rm H}^\star\tau_1$.

\subsection{PA Position Optimization for Special Case}
Next, given the updated WPCN resource variables $\tau_1^{(i)}$, $\tau_2^{(i)}$, $p_{\rm H}^{(i)}$, and $\mathbf{p}^{(i)}$, the PA position optimization block $\mathbf{x}^{(i)}$ is updated by solving
\begin{subequations}
	\begin{align}
		&(\mathbf{P_{2.2}}):
		\min_{\{\mathbf{x}, p_{\rm H}\}}
		P_{\rm H}^{\rm PAM}
		\sum\nolimits_{m=1}^{M}
		\frac{\Delta x_m}{v}+p_{\rm H}\tau_1^{(i)}
		\\
		&\mathrm{s.t.}~\tau_2^{(i)} \log_2\left(1+\sum\nolimits_{u=1}^{U}\frac{p_u^{(i)}|h_u|^2}{\sigma^2}
		\right)\geq U R_{\min}, \\
		&~~~~~p_u^{(i)}\tau_2^{(i)}\leq\tau_1^{(i)}\Phi_u,\ \forall u, \\
		&~~~~~(22{\rm e}), (22{\rm g}), (22{\rm h}), (22{\rm i}) \nonumber,
	\end{align}
\end{subequations}
where $p_{\rm H}$ is also optimized together with the PA positions, thereby providing an incentive for PAM through the potential reduction in WPT energy consumption. Similar to $(\mathbf{P_{1.2}})$, $(\mathbf{P_{2.2}})$ remains non-convex. Therefore, the fine-tuning-based SCA method developed for $(\mathbf{P_{1.2}})$ can be applied to $(\mathbf{P_{2.2}})$.

The AO-based algorithm is summarized in Algorithm 2. After convergence, the obtained PA positions are substituted into the original channel model, and the WPCN resource allocation is re-optimized to obtain the final energy consumption.

\vspace{-5pt}
\subsection{Convergence and Computational Complexity}
\subsubsection{Convergence}
The convergence of the proposed AO-based algorithm is analyzed through its two subproblems. Since $(\mathbf{P_{2.2}})$ is a minimization problem, the objective value of its convex approximation provides an upper bound that is monotonically tightened over SCA iterations for fixed WPCN resource allocation variables. Meanwhile, $(\mathbf{P_{2.1}})$ is optimally solved at each iteration. By alternating between the two subproblems, the objective value of $(\mathbf{P_2})$ is non-increasing and bounded from below, ensuring convergence to a finite limit.

\subsubsection{Computational Complexity}
The computational complexity of the proposed AO-based algorithm is determined by the computational costs of solving the two subproblems, i.e., WPCN resource allocation and PA position optimization. The former is solved by a one-dimensional search over $\tau_1$ with a computational complexity of $\mathcal{O}(N_{\tau_1})$, where $N_{\tau_1}$ denotes the number of candidate values of $\tau_1$.  Since the convex problem involves $(MU+M+1)$ variables, its computational complexity is $\mathcal{O}((MU+M+1)^3)$ per SCA iteration. The latter is solved by the SCA method, and its computational cost is about $\mathcal{O}(N_{\rm SCA}(MU+M+1)^3)$. Let $I_{\rm AO}$ denote the number of AO iterations. Hence, the total computational complexity is about $ \mathcal{O}\!\left[I_{\rm AO}\left(N_{\tau_1}+N_{\rm SCA}(MU+M+1)^3\right)\right] $.

\begin{algorithm}[t]
	\caption{AO-Based Algorithm}
	\begin{algorithmic}[1]
		\State \textbf{Initialize:} $\mathbf{x}^{(0)}$, $i\gets1$;
		
		\Repeat
		
		\State \textbf{\%WPCN Resource Allocation:}

		\State Update $\{\tau_1^{(i)},\tau_2^{(i)},p_{\rm H}^{(i)},\mathbf{p}^{(i)}\}$ solving $ (\mathbf{P_{2.1}}) $;
		
		\State \textbf{\%PA Position Optimization:}
		
		\State Update $ \{\mathbf{x}^{(i)}\} $ by solving $ (\mathbf{P_{2.2}}) $;
		
		\State $i\gets i+1$;
		
		\Until{The preset AO stopping condition is reached}
		
		\State \Return
		$\left\{
		\mathbf{x}^{\star},
		\tau_1^{\star},
		\tau_2^{\star},
		p_{\rm H}^{\star},
		\mathbf{p}^{\star}
		\right\}$
		
	\end{algorithmic}
\end{algorithm}

\section{Numerical Simulation}
In this section, we evaluate the achievable minimum energy consumption of the PA-assisted WPCN system with PAM costs. For ease of presentation, the general case and the special case are referred to as the adaptive-PAM-duration (APD) and fixed-PAM-duration (FPD) cases, respectively.

The simulation settings are described as follows, with reference to existing works, e.g., \cite{Y_Li_2026_1, Y_Xu_2026_2}. The $M$ PAs are initially and uniformly distributed along the waveguide with length $L_x$, while $ U $ IoTDs are randomly distributed on the ground within a rectangular area of size $L_x \times L_y$. The system parameters are set as listed in Table I. For each simulation setting, multiple independent IoTD deployments are generated and the corresponding performance results are averaged.

Several benchmark schemes are included to assess the effectiveness of the proposed algorithms and investigate the role of PAM costs in the design of the PA-assisted WPCN.

\begin{table}[t]
	\centering
	\caption{Simulation Parameters List}
	\begin{tabular}{ c c c }
		\hline
		Parameter & Notation & Value \\ 
		\hline
		Number of PAs & $M$ & 2-20 \\
		Number of IoTDs & $U$ & 1-8 \\
		Waveguide length & $L_x$ & 10 m \\
		Service area width & $L_y$ & 10 m \\
		Vertical distance & $d$ & 3 m \\
		Time block duration & $\tau$ & 1 s \\
		PAM duration & $\tau_0$ & 0.1 s \\
		PA moving speed & $v$ & 2 m/s \\
		Carrier frequency & $f_c$ & 28 GHz \\
		Effective refractive index & $n_{\rm eff}$ & 1.4 \\
		Minimum PA separation & $\Delta_{\min}$ & $\lambda/2$ \\
		Noise power & $\sigma^2$ & $10^{-15}$ W \\
		HAP maximum transmit power & $P_{\rm H}^{\max}$ & 10 W \\
		IoTD maximum transmit power & $P_u^{\max}$ & 0.1 W \\
		PAM power consumption & $P_{\rm H}^{\rm PAM}$ & 0.1-1 W \\
		Minimum throughput requirement & $R_{\min}$ & 0.1 - 0.9 bits/Hz \\
		Nonlinear EH circuit parameter & $\Phi_{\max}$ & 0.024 W \\
		Nonlinear EH circuit parameter & $a$ & 15 \\
		Nonlinear EH circuit parameter & $b$ & 0.014 \\
		\hline
	\end{tabular}
	\vspace{-10pt}
\end{table}

\subsubsection{Exhaustive Search (ES)-based Method}
It exhaustively searches the PA positions for the FPD case and both the PA positions and PAM duration for the APD case. The obtained results serve as the global-optimal benchmark for evaluating the optimality gap of the proposed algorithms \cite{M_Zeng_2026}.

\subsubsection{Fixed-PA (FPA) Method}
It keeps all PAs at their initial positions and optimizes the WPCN resource allocation. This method serves as a benchmark to evaluate the performance gain brought by PA position optimization.

\subsubsection{Movement-Costs-Ignoring (MCI) Method}
It ignores PAM costs during optimization and only accounts for them when evaluating the actual total energy consumption. Specifically, it ignores both PAM energy consumption and time overhead in the APD case and PAM energy consumption in the FPD case. After optimization, the corresponding PAM costs are included to calculate the actual total energy consumption. This method serves as a benchmark for evaluating the gain achieved by considering PAM costs in the optimization.

\vspace{-10pt}
\subsection{Convergence and optimality}
To evaluate the convergence and optimality of the proposed AO-based and ADMM-based algorithms, we plot the total energy consumption versus the number of iterations in Fig.~5. It is observed that the proposed algorithms converge rapidly, with the total energy consumption decreasing significantly during the first few iterations and then remaining nearly unchanged, indicating that stable solutions can be efficiently obtained. To further evaluate their optimality, we compare the converged solutions with the ES-based method. The ADMM-based algorithm in the APD case and the AO-based algorithm in the FPD case achieve optimality gaps of $0.94\%$ and $0.87\%$, respectively, confirming the near-optimal performance of the proposed algorithms.

\begin{figure}[!t]
	\centering
	\includegraphics[width=0.50\textwidth]{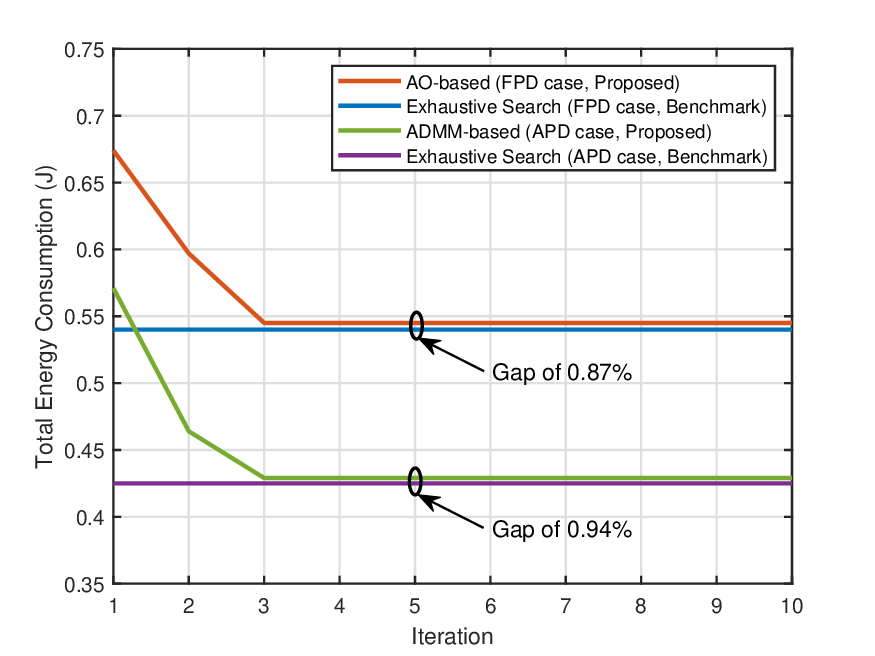}
	\caption{Total energy consumption versus iteration.}
	\label{Fig5}
	\vspace{-10pt}
\end{figure}

\vspace{-10pt}
\subsection{Impact of QoS Requirement on Total Energy Consumption}
Fig.~6 presents the total energy consumption versus the minimum QoS requirement $R_{\min}$.

First, the total energy consumption of all schemes increases with $R_{\min}$, as a higher QoS requirement requires a higher data transmission rate. Consequently, the PAs may need to move to more favorable positions, resulting in higher PAM energy consumption, while more energy is also required during the WPT phase, leading to higher WPT energy consumption. Moreover, due to the logarithmic relationship between achievable rate and received SNR, the total energy consumption exhibits an exponential-like increase with $R_{\min}$.

Second, the FPA method consumes more energy than the proposed algorithms, with the difference becoming more evident as $R_{\min}$ increases. This is because the proposed algorithms establish more favorable channels for WPT and WIT by jointly adjusting the PA positions, in terms of both the large-scale path loss and small-scale phase characteristics. This demonstrates that PAM can effectively reduce the system energy consumption, especially under high QoS requirements.

Third, the total energy consumption of the AO-based algorithm is higher than that of the ADMM-based algorithm. This can be attributed to the ability to adaptively optimize the PAM duration in the APD case and allocate the available time among PAM, WPT, and WIT, thereby achieving a better tradeoff between the PAM and WPCN. This demonstrates the benefit of adaptively optimizing the PAM duration in reducing the total system energy consumption.

\begin{figure}[!t]
	\centering
	\includegraphics[width=0.50\textwidth]{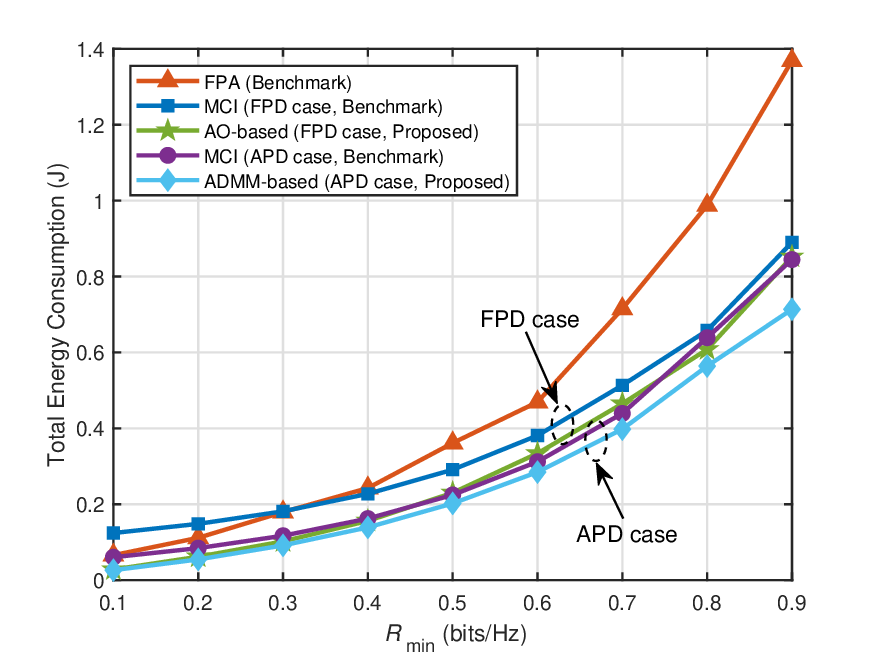}
	\caption{Total energy consumption versus minimum QoS requirement $R_{\min}$.}
	\label{Fig6}
	\vspace{-10pt}
\end{figure}

Finally, the MCI method consumes more energy than the proposed algorithms in both cases. This is because joint energy–time optimization jointly accounts for the energy and time costs of PAM, enabling the proposed algorithms to achieve a better tradeoff between PAM and WPCN. More importantly, as further quantified in Table~II, the energy-saving gains achieved by accounting for PAM costs increase as $R_{\min}$ decreases, reaching up to 77.51\% and 56.61\% in the FPD and APD cases, respectively. These results highlight the importance of accounting for PAM costs to achieve energy-efficient system operation, particularly under low QoS requirements.

\begin{table}[t]
	\centering
	\caption{Energy-Saving Gains Versus $R_{\min}$}
	\label{tab:energy_saving_gain}
	\setlength{\tabcolsep}{2pt}
	\begin{tabular}{c|cccccc}
		\hline
		Case & \multicolumn{6}{c}{$R_{\min}$ (bits/Hz)} \\
		\cline{2-7}
		& 0.1 & 0.2 & 0.3 & 0.4 & 0.5 & 0.6 \\
		\hline
		FPD & \textbf{77.51\%} & 58.43\% & 43.58\% & 31.38\% & 21.08\% & 12.35\% \\
		APD & \textbf{56.61\%} & 35.37\% & 21.77\% & 14.33\% & 9.96\% & 8.77\% \\
		\hline
	\end{tabular}
	\vspace{-10pt}
\end{table}

\begin{figure}[!t]
	\centering
	\includegraphics[width=0.50\textwidth]{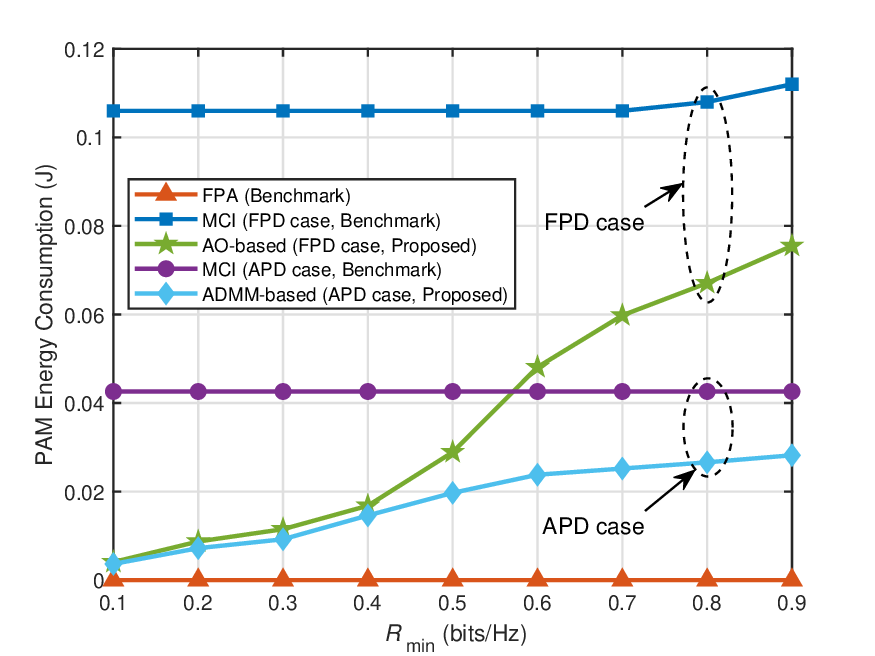}
	\caption{PAM energy consumption versus minimum QoS requirement $R_{\min}$.}
	\label{Fig7}
	\vspace{-10pt}
\end{figure}

\vspace{-10pt}
\subsection{Impact of QoS Requirement on PAM Energy Consumption}
Furthermore, Fig.~7 presents the PAM energy consumption versus the minimum QoS requirement $R_{\min}$, where the PAM energy consumption of both the AO- and ADMM-based algorithms is relatively low at low $R_{\min}$ and increases as $R_{\min}$ increases in both cases, as higher QoS requirements motivate the PAs to move to more favorable positions to get more favorable channel conditions. Meanwhile, the PAM energy consumption of the FPA and MCI methods remains unchanged, as the former fixes all PAs while the latter moves them directly to the most favorable positions regardless of PAM costs. Furthermore, compared with the MCI method, the proposed algorithms consistently incur lower PAM energy consumption in both cases, particularly under low QoS requirements. This indicates that, by incorporating PAM costs into the optimization, the proposed algorithms balance the performance benefits of PAM against its associated costs, rather than moving the PAs directly to the most favorable positions as in conventional designs that neglect PAM costs. This explains why the proposed algorithms achieve greater energy-saving gains under low QoS requirements, as shown in Table II: they avoid excessive PAM, thereby reducing unnecessary PAM energy consumption and time overhead.

\begin{figure}[!t]
	\centering
	\includegraphics[width=0.50\textwidth]{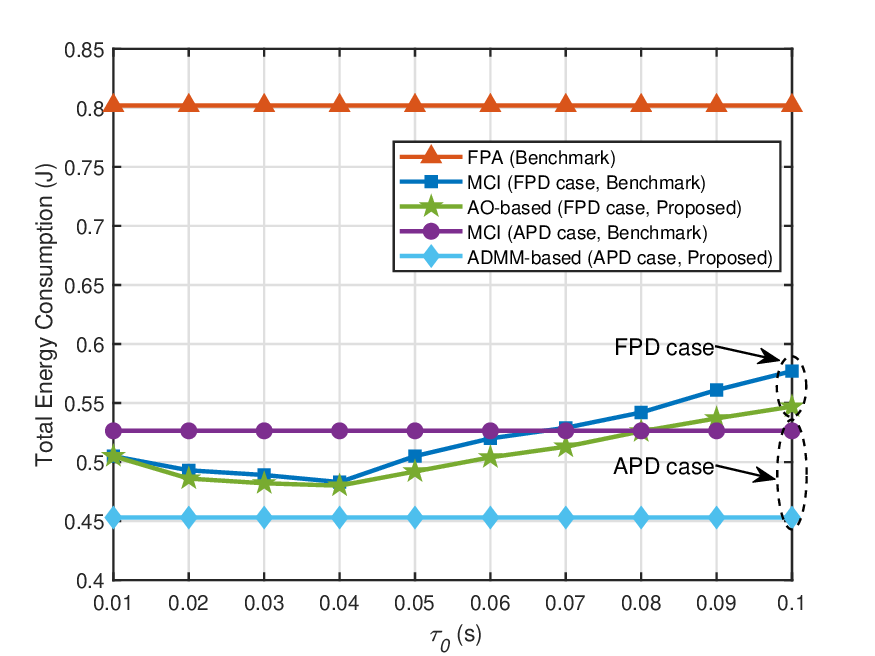}
	\caption{Total energy consumption versus PAM duration $\tau_0$.}
	\label{Fig8}
	\vspace{-10pt}
\end{figure}

\vspace{-10pt}
\subsection{Impact of PAM Duration on Total Energy Consumption}
Fig.~8 presents the total energy consumption versus the PAM duration $\tau_0$. In the FPD case, the energy consumption of the MCI and AO-based algorithm first decreases and then increases with $\tau_0$. When $\tau_0$ is too small, PAM is restricted, whereas an excessively large $\tau_0$ leaves less time for WPT and WIT, increasing the energy required to satisfy the QoS requirement. Therefore, there exists an optimal PAM duration that achieves a balance between PAM and WPCN operations. Meanwhile, the energy consumption of the FPA method and the MCI and ADMM-based algorithm in the APD case remains unchanged, as the former does not involve PAM and the latter adaptively optimizes $\tau_0$. Moreover, the ADMM-based algorithm in the APD case provides a lower bound on the minimum total energy consumption achievable by the AO-based algorithm in the FPD case. This result demonstrates the necessity of adaptively optimizing the PAM duration.

\begin{figure}[!t]
	\centering
	\includegraphics[width=0.50\textwidth]{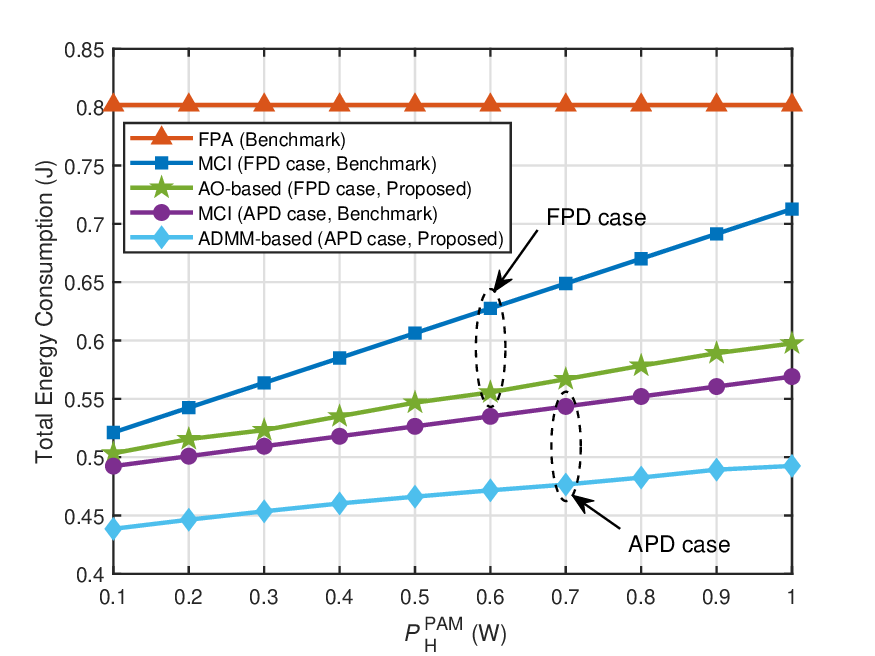}
	\caption{Total energy consumption versus PAM power $P_{\rm H}^{\rm PAM}$.}
	\label{Fig9}
	\vspace{-10pt}
\end{figure}

\vspace{-10pt}
\subsection{Impact of PAM Power on Total Energy Consumption}
Fig.~9 presents the total energy consumption versus the PAM power $P_{\rm H}^{\rm PAM}$. The FPA method remains unchanged because no PAM is performed, whereas the other methods exhibit increasing total energy consumption with increasing $P_{\rm H}^{\rm PAM}$, as higher PAM power directly increases the PAM energy consumption. In the FPD case, the gap between the AO-based algorithm and the MCI method widens as $P_{\rm H}^{\rm PAM}$ increases, indicating that considering PAM energy consumption becomes increasingly important when the PAM power is high. Among all methods,  the ADMM-based algorithm in the APD case exhibits the slowest energy increase, demonstrating that jointly considering the PAM duration $\tau_0$ and PAM energy consumption enables the system to better adapt to variations in PAM power, thereby achieving greater robustness against increases in PAM power.

\subsection{Impact of Number of PAs on Total Energy Consumption}
Fig.~10 presents the total energy consumption versus the number of PAs $M$. As $M$ increases, the total energy consumption of all methods decreases, with diminishing reductions. This is because more PAs provide additional spatial degrees of freedom for adjusting the wireless channels. However, as $ M $ further increases, the resulting benefits diminish. These results demonstrate that deploying more PAs can reduce the total energy consumption of PA-assisted WPCNs.

\begin{figure}[!t]
	\centering
	\includegraphics[width=0.50\textwidth]{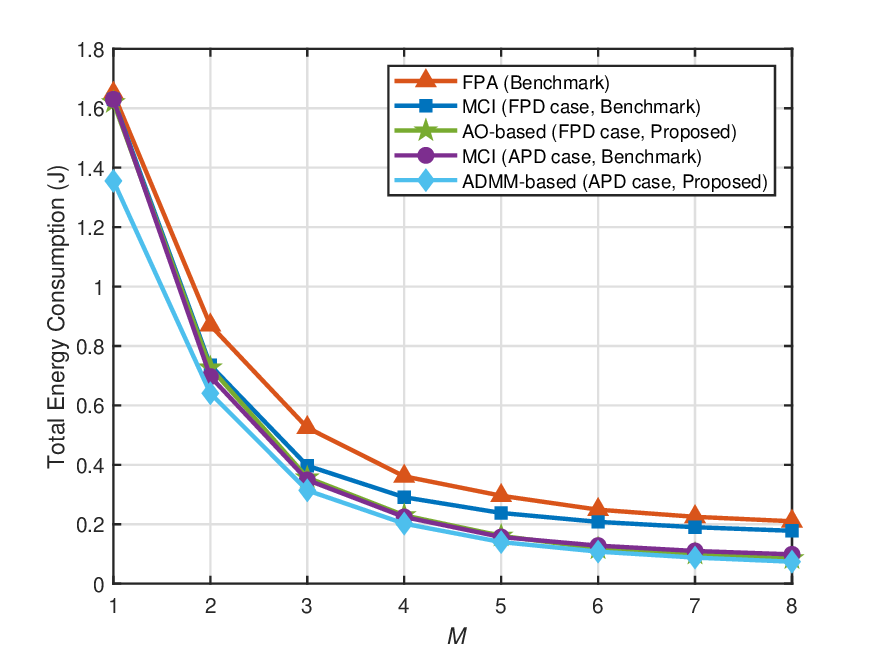}
	\caption{Total energy consumption versus number of PAs $M$.}
	\label{Fig10}
	\vspace{-10pt}
\end{figure}

\begin{figure}[!t]
	\centering
	\includegraphics[width=0.50\textwidth]{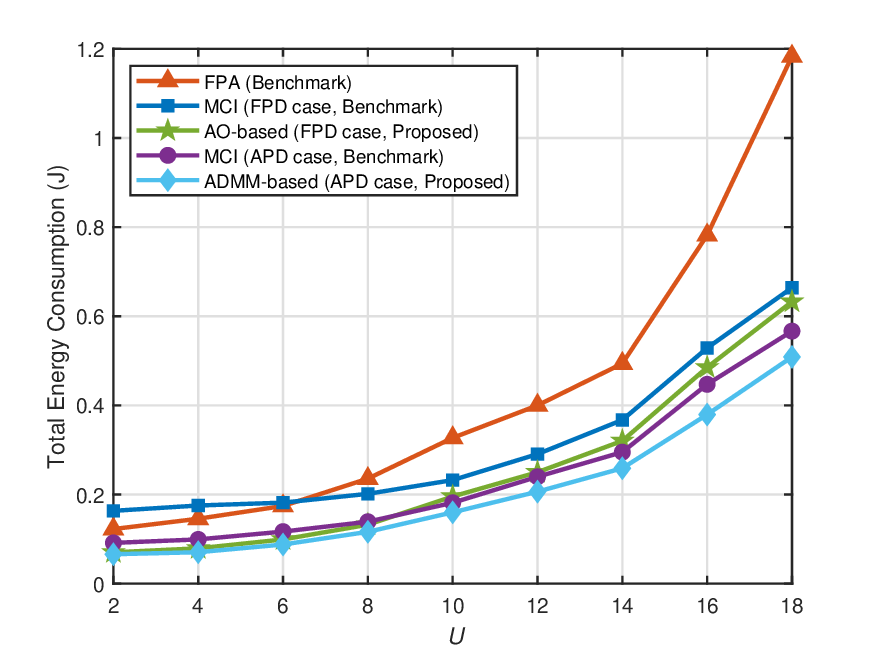}
	\caption{Total energy consumption versus number of IoTDs $U$.}
	\label{Fig11}
	\vspace{-10pt}
\end{figure}

\subsection{Impact of Number of IoTDs on Total Energy Consumption}
Fig.~11 presents the total energy consumption versus the number of IoTDs $U$. As $U$ increases, all methods consume more energy, with the growth becoming increasingly pronounced. This is because more IoTDs impose stricter aggregate QoS requirements, requiring more transmission and harvested energy. Due to the logarithmic relationship between achievable rate and received SNR, the total energy consumption increases rapidly with $U$, resulting in an exponential-like growth.

\vspace{-10pt}
\section{Conclusion}
This paper studied a PA-assisted WPCN with PAM costs and minimized the total system energy consumption by jointly optimizing the PA positions and WPCN resource allocation subject to the QoS requirements. For the general and special cases, the ADMM-based algorithm and lower-complexity AO-based algorithm were proposed, respectively. Numerical results demonstrated that the proposed algorithms achieved near-optimal performance. It is also concluded that accounting for both the energy consumption and time overhead of PAM enables the system to effectively balance the benefits of PA movement against its associated costs, thereby reducing unnecessary PA movement and the total system energy consumption.

\end{document}